\documentclass[10pt]{elsarticle}
\pdfoutput=1
\usepackage{color}
\usepackage{slashed}
\usepackage{amsmath}
\usepackage{amssymb}
\usepackage{graphicx}

\newcommand{\beqn}{\begin{eqnarray}}
\newcommand{\eeqn}{\end{eqnarray}}

\definecolor{purple}{rgb}{0.8,0,0.6}
\newcommand{\revision}[1]{{#1}}

\begin{document}

\title{Momentum space topology of QCD}

\author[ARIEL]{M.A.~Zubkov
\footnote{
Corresponding author,
e-mail: zubkov@itep.ru}\footnote{on leave of absence from Institute for Theoretical and Experimental Physics, B. Cheremushkinskaya 25, Moscow, 117259, Russia} }

\address[ARIEL]{Physics Department, Ariel University, Ariel 40700, Israel}

\begin{abstract}
We discuss the possibility to consider quark matter as the topological material. We consider hadronic phase (HP),  the  quark - gluon plasma phase (QGP), and the hypothetical  color - flavor locking (CFL) phase. In those phases we identify the relevant topological invariants in momentum space. The formalism is developed, which relates those invariants and massless fermions that reside on vortices and at the interphases. This formalism is illustrated by the example of vortices in the CFL phase.
\end{abstract}

\maketitle

\section{Introduction}

QCD has the reach phase structure, which has not been yet investigated exhaustively. Various external conditions (temperature, baryonic chemical potential, external magnetic field, pressure, isospin chemical potential, chiral chemical potential, etc) cause the appearance of different phases (for the review see \cite{QCDphases,1,2,3,4,5,6,7,8,9,10,CFL0,CFL,m1,m1m8,Nitta,latticeQCD,diquarks,Zhitnitsky} and references therein). Some of those phases are connected continuously, while the others are separated by the true phase transitions. The major known phases have different symmetries. At the same time, within each phase the topological phase transitions \cite{Volovik2003} are possible, which are characterized by the same symmetry but different values of momentum space topological invariants. While the breakdown of various symmetries across the transitions between the phases has been discussed in details in many cases, the consideration of the possibility for the topological phase transitions is still in its infancy.

Momentum space topology (i.e. the theory of topological invariants in momentum space) has been widely discussed recently in the context of relativistic theories (see \cite{So1985,IshikawaMatsuyama1986,Kaplan1992,Golterman1993,Volovik2003,Horava2005,Creutz2008,Kaplan2011,Creutz2011,ZubkovVolovik2012,Zubkov2012,Volovik2000,Volovik2010,VZ2016,Z2016,Z2016AOP}).
Originally momentum space topology was developed in the condensed matter theory
   \cite{HasanKane2010,Xiao-LiangQi2011,Schnyder2008,Kitaev2009,Grinevich1988,BevanNature1997}, where it is nowadays used widely for the description of various systems among which the most well known are the topological insulators. The first step towards the prediction of the appearance of the new topological sub - phases within each phase of QCD is the identification of the relevant topological invariant.

The different sub - phases of the given phase would correspond to different values of a topological invariant. The continuous deformation of the system between the states with different values of the topological invariant is impossible. The only possibility to move between such phases is to cross a phase transition, which is  accompanied by the appearance of the singularity of the fermionic Green function, and the change of the value of the topological invariant composed of this Green function. Typically, such a topological phase transition is accompanied by the discontinuities in thermodynamical quantities.

The present paper is devoted to the identification of the relevant topological invariants in the following phases of QCD:

\begin{enumerate}

\item{} Hadronic phase (HP). This is the most well - known phase of QCD, which corresponds to the observed hadronic and nuclear matter  \cite{QCDphases,1,2,3,4,5,6,7}. The latter corresponds to the nonzero chemical potential, which is typical for the matter inside the nuclei. There are the two sub - phases of the HP, which are separated by the vapor - gas first order phase transition, having its endpoint, so that those two sub - phases are connected continuously.

\item{} Quark - gluon plasma phase (QGP). It appears at the temperatures above the confinement - deconfinement transition  \cite{QCDphases,1,2,3,4,5,6,7,8,9,10}. Notice, that recent data of lattice numerical simulations \cite{latticeQCD} indicate, that at the realistic values of quark masses this transition may appear to be a crossover, and that the low temperature Hadronic phase is connected continuously with  the QGP phase. That means, that those two phases actually have the same symmetry. In the limit of massless $u$ and $d$ quarks however, those two phases are indeed different and the transition between them corresponds to the breakdown of chiral symmetry.

\item{} When the quark chemical potential is increased, the hadronic phase undergoes first the gas - liquid phase transition, after that the other transitions are possible \cite{QCDphases,1,2,3,4,5,6,7}.  It is widely believed, that the further increase of chemical potential finally leads to the formation of color superconductivity \cite{CFL0,CFL,m1,m1m8,Nitta}. According to this hypothesis at extremely high values of chemical potential the  CFL phase appears, which is characterized by the condensation of the quark - quark pairs composed of $u$, $d $ and $s$ quarks.

\end{enumerate}

\begin{figure}[!thb]
\begin{center}
\includegraphics[height=60mm,width=80mm,angle=0]{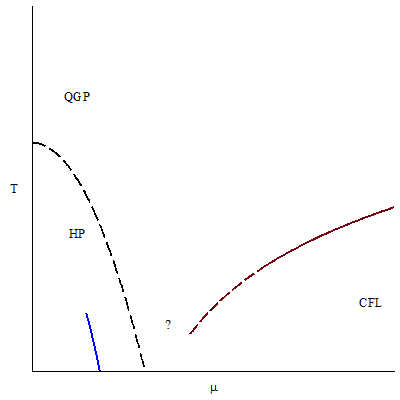}
\end{center}
\caption{\label{fig1}  The expected phase diagram of QCD at finite baryon chemical potential $\mu$ and finite temperature $T$ is represented schematically (assuming the validity of the hypothesis that at large values of $\mu$ the color superconductivity appears).}
\end{figure}

Those three phases are represented in Fig. \ref{fig1}, where the expected phase diagram is given schematically.
The hypothetical existence of the topological phase transitions within various phases of QCD follows the analogy with the topological insulators  \cite{HasanKane2010,Xiao-LiangQi2011,Schnyder2008,Kitaev2009}, which have reach topological structure caused by the similar topological invariants. But this is not the only consequence of the existence of the momentum space topological invariants. The analogy to the condensed matter physics also prompts that another consequence would be related to the existence of the specific massless boundary states.  Recall, that in condensed matter physics this is the massless boundary state, which identifies  bulk topological matter (either  topological semimetal \cite{Soluyanov2015,SchnyderBrydon2015,Mizushima2016,Bansil2016,Yonezawa2016} or a topological insulator \cite{Witten2015,Wang2014,Kapustin2015,Kitaev}). In the same way, in QCD the existence of nonzero value of the topological invariant might be related to the existence of massless fermions at the boundary of the region of space filled by the quark matter of the given state.  To find the massless boundary states using the topological invariant in momentum space through the bulk - boundary correspondence \cite{Volovik2003,Gurarie2011, EssinGurarie2011,Volovik2009,Silaev} we need the finite regions of space filled by the given state of quark matter, outside of these regions the state of matter is different while the symmetry is the same. In this case one may use the same topological invariant at both sides of the boundary.

The appearance of the finite region of space filled by the given state of quark matter is typical, for example, for the heavy ion collisions \cite{heavyions}, or inside the stars \cite{stars}. Out of such regions of space the quark matter is typically in the state with different symmetry. This does not allow to use directly the topological invariants protected by symmetry  for the description of the fermion zero modes at the interphases between the phases of different symmetries. Such a description would become possible if the topological phase transition
is found inside a certain phase so that the two sub - phases with the same symmetry but different values of the topological invariant may coexist. Then at the interphase between them one would find the massless fermions through the bulk - boundary correspondence.

The power of the topological invariants in momentum space is demonstrated below through the consideration of vortices  that are formed inside the CFL phase. (For the recent alternative discussion of the possible topological properties of the color superconductivity see \cite{Nishida:2010wr}.) Then the bulk topological invariant is intimately related to the zero modes existing at the topological defects. For the previous discussion of the fermion zero modes on those vortices see \cite{nitta_1,nitta_2}. Using the ideas of the representation of such a correspondence  \cite{Silaev} (see also \cite{EssinGurarie2011}) in terms of the mixed $r-p$ invariant (the invariant in mixed coordinate - momentum space), we apply the methodology of \cite{Z2016AOP} that operates with the Wigner transformation \cite{Wigner,star2,Weyl,berezin} of the Green function and describe the fermion zero modes that reside on vortices in the CFL phase.

Thus, in the present paper the investigation of the QCD phase structure is proposed based on the consideration of momentum space topology. In order to point out the possible place of this methodology in the whole pattern let us recall the basic features of QCD. The important feature of QCD (which makes it essentially different, say, from Quantum Electrodynamics) is that the complete analytical treatment of the phase structure is impossible because of the relevance of the nonperturbative effects. The only sufficient method of investigation is lattice numerical simulations, which are technically rather complicated. The commonly accepted way of the investigation of the nonperturbative aspects of QCD (including the phase structure) consists of the combination of:

\begin{enumerate}

\item{Precise analytical methods.}  These methods include, first of all, the ordinary perturbation theory applied to the calculation of Green functions at large momentum transfer. Also those methods include: consideration of classical solutions, including instantons; the technique of various Abelian projections and classification of the topoogical defects existing in those projections \cite{Polykarp} (those topological defects, – monopoles and vortices, - are the special classes of the configurations of the gauge field).

 The precise perturbative methods allow to describe more or less exhaustively only the limited class of phenomena (deep inelastic scattering, for example). The other two mentioned above methods are typically used in order to construct the phenomenological models and to check them by numerical simulations. An example is the combined study of the dual superconductor mechanism of confinement. In this mechanism the quark – antiquark string appears as the Abrikosov vortex in the effective model, in which the monopoles of Maximal Abelian Projection are condensed and play the role of the condensed scalar field of the effective Abelian Higgs Model \cite{Polykarp}.

\item{Construction and investigation of approximate phenomenological models.} As it was mentioned above, to these models belong, in particular, the dual superconductor model, that is a certain Abelian Higgs model, in which the scalar field represents the field of the Abelian monopoles. Those monopoles are identified as configurations of the non – Abelian gauge field considered in the so - called Maximal Abelian Projection.

Another example of the phenomenological model of QCD is given by the class of Nambu $-$ Jona $–$ Lasinio models \cite{NJL}, where the nonperturbative interactions between quarks are approximated by the effective four $–$ fermion interaction. Those models appear to be very successful for the description of chiral symmetry breaking and the calculation of spectrum of light scalar mesons. On the other hand, in their classical form they are not able to explain confinement.

It is also worth mentioning here that various models were proposed that approximate vacuum of QCD by an instanton liquid \cite{liquid}.

The methods based on the partial resummation of diagrams combined with the exact equations (like the Schwinger – Dyson ones) also may be considered as the approximate phenomenological models. Such methods, being approximate, nevertheless may be useful in the partial treatment of the non – perturbative phenomena.

\item{Numerical simulations.}
Numerical simulations represent the most powerful method of the investigation of non $–$ perturbative features of QCD. Their application is limited, in principle, by the power of supercomputers only. Besides, the relevant numerical algorithms are still not developed, say, for the case of the non $–$ zero baryon chemical potential \cite{Born}. As a result the phase structure of QCD is understood much better at $\mu = 0$, and nonzero $T$ than for $T = 0$ and nonzero $\mu$.

\end{enumerate}

Usually the first step in this list precedes the third step. It may be, however, combined with the second one. The work on the numerical simulations requires much more resources than the preceding two steps. Therefore, it represents typically the final step of the study. In the present paper the completely new methodology of the investigation of QCD is proposed. It starts from the analytical treatment that, according to the above classification, belongs to the step 1. It consists of the identification of the topological invariants in momentum space, and their relation to various observable phenomena. The values of the proposed invariants are calculated for the noninteracting quarks (when the exchange by gluons is turned off) with the constant mass matrix of the form specific for the phase under consideration. Such a system of noninteracting quarks may be connected continuously to the system of interacting fermions at the given external conditions. Then the value of the given topological invariant will be the same for those two systems. The noninteracting system may also be connected via a phase transition to the interacting one. Then the values of the topological invariants may be different.

The very first step of an investigation presented in the given paper is not accompanied by numerical simulations, or by the construction of phenomenological approximate models. Unfortunately, without the numerical simulations this is not possible as a matter of principle to answer the question on the presence or absence of the phase transition caused by the exchange by gluons. Therefore, we do not know does the gluon exchange change the value of the particular topological invariant compared to that of the model with noninteracting quarks. Only the future numerical lattice investigations will give the answer. This should constitute the separate study and is out of the scope of the present paper.

The paper is organized as follows. In Section \ref{SymmetricPhase} the topological invariants specific for the QGP phase are considered. In Sect. \ref{Parity} the topological invariants relevant for the description of HP are discussed. In Sect. \ref{DomainWall} the hypothetical domain walls in QCD are discussed based on the methodology of bulk - boundary correspondence. This correspondence allows to calculate the number of massless fermions on the domain walls.   In Sect. \ref{SectCFL} the CFL phase is discussed. Further, in Sect. \ref{SectVORT} the methodology that allows to identify massless fermions on the vortices is developed. This methodology is applied in Sect. \ref{SectCFLVORT} to vortices existing in the CFL phase. In Sect. \ref{SectGen} the common topological invariant for all gapped phases of QCD is proposed. In Sect. \ref{SectConcl} we end with the conclusions.

\section{Quark - gluon plasma (QGP) phase}
\label{SymmetricPhase}

The discussion of this section is to a certain extent similar to the consideration of the symmetric high temperature phase of the whole Standard Model presented in the Second section of \cite{VZ2016}. For vanishing quark chemical potential the quark gluon plasma phase appears at the temperatures above the confinement - deconfinement phase transition. In the approximation when the $u$ and $d$ quarks are considered as massless while the $s$ quark mass is kept equal to its physical value close to $100$ MeV the transition between the two phases is the true phase transition, which is accompanied by the restoration of the chiral $SU(2)$ symmetry. The chiral $U(1)$ symmetry remains broken due to the instantons, but its subgroup $Z_2$ is already not broken. This subgroup corresponds to the transformations $q=(u,d,s) \rightarrow e^{i\pi \Pi_{u,d} \gamma^5/2} q$, where $\Pi_{u,d}$ is the projector to the states of $u$ and $d $ quarks:
$$
u \rightarrow e^{i\pi \gamma^5/2}\,u, \quad d \rightarrow e^{i\pi \gamma^5/2}\,d, \quad s \rightarrow s
$$
This is because the corresponding transformation of measure gives $ D\bar{u}Du \to e^{\pi\,i\, \nu}D\bar{u}Du$ and $ D\bar{d}Dd\to e^{\pi\,i\, \nu}D\bar{d}Dd$, and therefore, $$ D\bar{u}DuD\bar{d}Dd\to e^{2\pi\,i\, \nu}D\bar{u}DuD\bar{d}Dd$$
Here $\nu$ is the integer number of $SU(3)$ instantons.

First of all, let us consider the following topological invariant expressed through the fermion two point Green function ${\cal G}$ determined in the 4 dimensional momentum-frequency space \revision{\cite{Grinevich1988,Volovik2003} }:
\begin{equation}
{\cal N}_3 = {\bf tr} ~{\cal N}~~,~~{\cal N}  =
\frac{1}{{24\pi^2}}e_{\mu\nu\lambda\gamma}~
 ~\int_{\sigma}~  dS^{\gamma}
~ {\cal G}\partial_{p_\mu} {\cal G}^{-1}
{\cal G}\partial_{p_\nu} {\cal G}^{-1} {\cal G}\partial_{p_\lambda}  {\cal
G}^{-1}~.
\label{TopInvariantMatrix}
\end{equation}
The integral here is over the $S^3$ surface
$\sigma$ embracing the point in the 4D space ${\bf p}=0, p_4=0$, where $p_4$ is the  frequency
along the imaginary axis; ${\bf tr}$ is the trace over the fermionic indices.
 We imply the fixing of Landau gauge, in which the functional $\int d^4x {\rm Tr}\, A^2$ is minimized with respect to local gauge transformations, where $A \in su(3) $ is the gauge field.

 The topological invariant Eq. (\ref{TopInvariantMatrix}) given by the integral over the closed $3D$ surface in momentum space is well defined only if the values of imaginary frequency are continuous. At finite temperatures they are discrete. In order to define such an invariant we need the analytical continuation of the Green function to the continuous values $p_4$ of imaginary frequency. In this case the invariant protects the pole or zero at $p_4 = 0$ (if any) of the resulting analytically continued fermion propagator. At zero temperature this pole appears when the system is gapless. At finite temperature the discrete Matsubara frequency never vanishes, and the physical propagator depending on it is not singular even if the analytically continued propagator is singular at $p_4=0$.

In the absence of the exchange by the $SU(3)$ gauge bosons for the right handed Weyl fermions representing $u$ and $d$ quarks
one has ${\cal N}_3=1$, and ${\cal N}_3=-1$ for the left handed Weyl fermions. Both are gapless in the absence of interactions in the chiral limit.  At the same time for the massive $s$ quarks the Fermi point is absent, and the corresponding components of the Green function give vanishing contribution to ${\cal N}_3$. Therefore, altogether the value of ${\cal N}_3$ is vanishing.

At the same time due to the mentioned above $Z_2$ symmetry the following expression gives the value of the topological invariant protected by symmetry
\begin{equation}
{\cal N}_3^{Z_2}= {\bf tr} \, \left[ e^{i\frac{\pi}{2}\Big( \Pi_{u,d} \gamma^5 + 1\Big)} {\cal N}
\right] \,
\label{GeneratingFunctionWYC}
\end{equation}
Again, suppose first that there are no interactions due to the color gauge bosons. Smooth continuous transformations of the system cannot change the value of a topological invariant.
If the region of the phase diagram where chiral symmetry is not broken, and which is connected continuously to the system of noninteracting $u,d,s$ quarks, we may substitute to Eq. (\ref{GeneratingFunctionWYC})
\begin{equation}
{\cal G}^{-1}\sim p\gamma, \, p^2\to 0
\end{equation}
This gives ${\cal N}_3^{Z_2} =2 N_f N_c = 12 $. (In this case the number of quark flavors is $N_f=2$.) For the method of calculation see, for example, \cite{Z2016AOP}. Below we represent the similar calculation for the more complicated case of interacting fermions.

The so – called thermal mass appears in QED at high temperatures \cite{LeBellac}, which may be demonstrated using the technique of HTL (the hard thermal loop). This thermal mass appears precisely in the mentioned above analytically continued fermion propagator, and it actually means the absence of the pole in it. The similar behavior of the fermion propagator has been also predicted in the framework of perturbative QCD at finite temperatures \cite{LeBellac}. Namely, the one loop results give the fermion propagator
\begin{equation}
{\cal G}({\bf p}, p_4) = \frac{1}{2}(\gamma^0 - \gamma \hat{\bf p})\Delta^{-1}_-(p_4,{\bf p}) + \frac{1}{2}(\gamma^0 + \gamma \hat{\bf p})\Delta^{-1}_+(p_4,{\bf p})\label{Gnondef}
\end{equation}
Here $\gamma^k$ are the usual Dirac matrices (in Minkowski spacetime), $\hat{\bf p} = {\bf p}/|{\bf p}|$ while
\begin{equation}
\Delta_\pm = i p_4 \mp p - \frac{m_f^2}{2 p}\Big[(1\mp\frac{i p_4}{p})\,{\rm log}\,\frac{i p_4 + p}{i p_4 - p}\pm 2\Big]
\end{equation}
with $p = |{\bf p}|$ and the thermal mass $$m^2_f = \frac{1}{8}g^2_{SU(3)}C_F(T^2 + \frac{\mu^2}{\pi^2})$$
 Here $g_{SU(3)}$ is the QCD coupling constant, while $C_F = 4/3$.

One might naively think that for such a system the topological invariant similar to that of Eq. (\ref{GeneratingFunctionWYC}) should be equal to zero. However, we will see below, that this is not so because instead of the pole the zero of the Green function appears. The zeros of the Green functions represent an unusual phenomenon, which is not yet investigated in details. For the discussion of this issue see, for example, \cite{Zubkov:2012nm} and references therein.

We obtain at $p_4 \ll  p \ll m_f$:
\begin{eqnarray}
{\cal G}({\bf p}, 0)& = &  \frac{1}{2}(\gamma^0 - \gamma \hat{\bf p})\frac{p}{p^2 + m_f^2 (1-i\pi/2)} - \frac{1}{2}(\gamma^0 + \gamma \hat{\bf p})\frac{p}{p^2 + m_f^2 (1+i\pi/2)}\nonumber\\
& = &  \frac{1}{2}\frac{\gamma^0 p - \gamma {\bf p}}{p^2 + m_f^2 (1-i\pi/2)} - \frac{1}{2}\frac{\gamma^0 p + \gamma {\bf p}}{p^2 + m_f^2 (1+i\pi/2)}\nonumber\\
& \approx & \frac{ i \frac{\pi}{2}\gamma^0 p - \gamma {\bf p}}{ m_f^2 (1+\pi^2/4)}
\end{eqnarray}
In the opposite limit $p \ll p_4 \ll m_f$:
\begin{eqnarray}
{\cal G}(0,p_4)& = & \frac{- i \gamma^0 p_4}{ m_f^2 + p_4^2}\approx  \frac{- i \gamma^0 p_4}{ m_f^2 }\label{Gdef}
\end{eqnarray}
For the intermediate values of $p,p_4$ the Green function neither has the zeros nor the poles. One can see, that the presented form of the propagator possesses an interesting property. It has no poles at finite temperature due to the appearance of the thermal mass. However, instead of the pole the zero appears at $p=p_4=0$. In a small vicinity of this point in momentum space we may take the integration surface in Eqs. (\ref{TopInvariantMatrix}), (\ref{GeneratingFunctionWYC}) in the form of the small sphere surrounding the point $p_4 = p = 0$ that crosses the axis $p=0$ at $p_4= \pm \epsilon$. Because we deal with the topological invariant, the form of the Green function may be deformed smoothly in such a way, that the singularities are not encountered. Thus instead of the discussed above form of the Green function we use for the calculation
\begin{eqnarray}
\tilde{\cal G}({\bf p}, p_4)& = & \frac{ i \gamma^0 (p\frac{\pi}{2}-p_4(1+\pi^2/4)) - \gamma {\bf p}}{ m_f^2 (1+\pi^2/4)}
\end{eqnarray}
Let us denote $g_a(p) = p_a$ for $a=1,2,3$ and $g_4(p) = - p\frac{\pi}{2}+p_4(1+\pi^2/4)$. Also let us introduce the notation $\hat{g}_a = \frac{g_a}{\sqrt{g_bg_b}}$. Then using the formalism developed in \cite{Z2016AOP} we obtain:
\begin{equation}
{\cal N}_3^{Z_2}= \frac{2 N_f N_c}{{12\pi^2}}\epsilon^{abcd} \int_{\sigma}~  \hat{g}_a d \hat{g}_b \wedge d\hat{g}_c \wedge d\hat{g}_d
\label{N3Z2C}
\end{equation}
Next, we introduce the parametrization
\begin{equation}
\hat{g}_4 = {\rm sin}\,\alpha, \quad \hat{g}_i = k_i\,{\rm cos}\,\alpha
\end{equation}
where $i=1,2,3$ while $\sum_{i}k^2=1$, and $\alpha \in [-\pi/2,\pi/2]$. This gives
  \begin{eqnarray}
\frac{{\cal N}^{Z^2}_3}{2 N_c N_f} &=&  \frac{1}{4 \pi^2} \epsilon^{ijk}\, \int_{\cal M} \, {\rm cos}^2 \alpha\, k_i\,  d\,\alpha  \wedge d k_j \wedge d k_k\nonumber\\ &=&  \frac{1}{4 \pi^2} \epsilon^{ijk}\, \int_{\cal M} \, k_i\,  d(\alpha/2+\frac{1}{4}{\rm sin}\,2\alpha)  \wedge d k_j \wedge d k_k \nonumber\\ &=& -\sum_l \frac{1}{4 \pi^2} \epsilon^{ijk}\, \int_{\partial{\Omega(y_l)}} \, k_i\,  (\alpha/2+\frac{1}{4}{\rm sin}\,2\alpha)   d k_j \wedge d k_k
\end{eqnarray}
In the last row $\Omega(y_l)$ is the small vicinity of point $y_l$ of momentum space, where vector $k_i$ is undefined. The absence of the singularities of $\hat{g}_k$ implies, that $\alpha \rightarrow \pm\pi/2$ at such points.

This gives
  \begin{eqnarray}
\frac{{\cal N}^{Z^2}_3}{2 N_c N_f} &=&   -\frac{1}{2}\sum_l \, {\rm sign}(g_4(y_l)) \,{\rm Res}\,(y_l)
\end{eqnarray}
As in \cite{Z2016AOP} we use the notation:
\begin{eqnarray}
{\rm Res}\,(y) &=&  \frac{1}{8 \pi} \epsilon^{ijk}\, \int_{\partial \Omega(y)} \, \hat{g}_i  d \hat{g}_j \wedge d \hat{g}_k
\end{eqnarray}
It is worth mentioning, that this symbol obeys $\sum_l {\rm Res}\,(y_i)=0$.
In the particular case of the deformed Green function of the form of Eq. (\ref{Gdef}) there are two points on the sphere surrounding the point $p_4 = 0, p=0$, where $p=0$. Those two points are $p_4 = \pm \epsilon, p=0$. As a result we obtain
$${\cal N}_3^{Z_2} =2 N_f N_c = 12 $$
The same answer is valid for the Green function of the form of Eq. (\ref{Gnondef}) because it is related to Eq. (\ref{Gdef}) by continuous deformation.
If the nonperturbative effects do not alter this prediction, then the value of the topological invariant of Eq. (1) equals to $2 N_f N_c = 12$ in the QGP phase at high temperatures as without interactions.

Thus we have an interesting situation: without interactions there is the pole of the Green function that corresponds to massless excitations. When interactions are taken into account at high temperature the pole of the Green function disappears. Instead the zero of the Green function appears. Although the topological invariant ${\cal N}_3^{Z_2}$ has the same values in those two states, it is clear, that they are essentially different. The transition between the two states should be accompanied by the transformation of the singularity of the Green function to its zero. At zero temperature this may occur through the phase transition only. However, at the finite temperature this is not so. The pole of the analytically continued Green function does not represent the physical singularity because the physical propagator (entering expressions for the thermodynamical quantities) is not singular at the discrete values of Matsubara frequencies.

\section{Hadronic phase (HP)}
\label{Parity}

As it was mentioned above, the hadronic phase is separated from the QGP phase if masses of $u$ and $d$ quarks are disregarded. According to the so - called Columbia plot \cite{deForcrand:2017cgb} the transition between the confinement and deconfinement phases is of the first order for sufficiently small $m_u, m_d, m_s$, and becomes a crossover at large enough values of masses (unfortunately, the present data obtained by different lattice regularizations contradict each other on the corresponding critical values of masses). Nevertheless, even if the real values of the current masses $m_u, m_d, m_s$ belong to the lower left corner of the Columbia plot, where the transition is of the first order, we may deform the theory smoothly at $T < T_c$ without a phase transition increasing the values of masses. Next, we come to the high temperature deconfinement phase via a crossover, and decrease the values of masses to original values.  Thus if nonzero current masses are taken into account, then the two phases are connected continuously.  Actually, taking into account those small but finite current masses we may consider these two phases as the one phase. The relevant topological invariant is the one given by
\cite{Volovik2009}
\begin{equation}
{\cal N}^K_3 =
\frac{1}{{24\pi^2}}e_{\mu\nu\lambda} \, {\bf tr} \, \left[ K
 ~\int  d^3p
~ {\cal G}\partial_{p_\mu} {\cal G}^{-1} \,
{\cal G}\partial_{p_\nu} {\cal G}^{-1} \, {\cal G}\partial_{p_\lambda}  {\cal
G}^{-1} \right]\,.
\label{TopInvariantMassive}
\end{equation}
Here $p_4=0$; the integral is over 3-momentum space;
and $K$ is the proper symmetry operation (it should either commute of anti - commute with $\cal G$).
As above, at finite temperatures we need to continue the Green function to the vanishing value of imaginary frequency $\omega= -ip_0 = p_4$.

Parity is not broken by QCD with $u,d,s$ quarks. Therefore, for vanishing chemical potential we take following \cite{VZ2016} $K$ as the composition of $C$ and $T$ transformations. At zero temperature and vanishing chemical potential the Green's function has the form
 \begin{equation}
{\cal G}(p)=   \frac{1}{A(-p^2)\gamma^\mu p_\mu- B(-p^2)}\,,
\label{eq:S(p)}
\end{equation}
where $\gamma^\mu$ are the Dirac matrices in Minkowski space, $p^2=-{\bf p}^2+p_0^2$, while $A$ and $B$ are diagonal hermitian matrices. The fermion mass matrix $m$ is given by the solution of equation
  $$A(-m^2) m = B(-m^2)$$
Here it is the matrix  $K=i\gamma^5\gamma^0$, which commutes with the Green's function at $p_4=\omega=-i p_0 = 0$. This is the matrix of the combination of  C and T.  If we take into account only three light quarks, then
 \begin{equation}
{\cal N}_3^{CT}=  N_c \times N_f ,
\label{eq:8g}
\end{equation}
 if  vacuum is connected continuously to the vacuum of massive noninteracting Dirac fermions. Here $N_c = 3$ is the number of colors while $N_f = 3$ is the number of flavors. To demonstrate this we take the fermion Green function of the form
 \begin{equation}
 G^{-1} =  \gamma^\mu p_\mu  -  m
\end{equation}
and obtain
\begin{eqnarray}
{\cal N}^{CT}_3 &=& i{e_{ijk}\over{24\pi^2}} ~
{\bf tr}\left[  \int_{\omega=0}   \frac{d^3p}{(\vec{p}^2+m^2)^2} ~\gamma^0\gamma^5 m  \gamma^i\gamma^j\gamma^k
~ \right]\nonumber\\
&=&  N_c \times N_f
\end{eqnarray}
Each fermion mode (enumerated by flavor and color) contributes $1$ to this quantity.

As it was mentioned above, according to the lattice numerical data the hadronic phase is connected continuously  to the QGP phase. The latter, in turn, is assumed to be connected continuously with the system of noninteracting fermions (this statement is to be verified by direct calculations, though). At finite values of chemical potential the form of the Green function is more complicated than Eq. (\ref{eq:S(p)}). However, the invariance under the combination of C and T is still present, and therefore, $K$ still commutes with the Green function at $\omega = 0$.

It is worth mentioning, that if we would consider two flavors of light quarks instead of three then the different values of ${\cal N}_3^{CT}$  may mark the same topological classes. This is because $B(-p^2)$ is defined up to the global $Z_2$ transformation $B(-p^2) \to B(-p^2) e^{i \pi \tau^3}$, where $\tau^3={\rm diag}(1,-1)$. The values of the ${\cal N}_3^{CT}$ with opposite signs mark the same topological classes, and the actual classification in this case is $Z/Z_2$ rather than $Z$. However, for the three quark flavors this is not so, and the opposite values of ${\cal N}_3^{CT}$ do represent different topological classes of vacua.

We may consider the phase diagram of QCD with the axes: temperature, baryonic chemical potential, external magnetic field, external electric potential, angular velocity of rotation, etc. Thus we describe quark matter that may be hot, dense, may experience the external magnetic field and external electric potential, may be rotated. Such states might appear both in the heavy ion collisions and inside the stars. We may also introduce the isotopic and strangeness chemical potentials, which are conserved in heavy ion collisions. This enriches further the phase diagram. In addition, we may consider the smooth deformation of the theory turning off slowly interactions between gauge fields and quarks. This adds an extra axis to the phase diagram. Although certain efforts we made during the last year to investigate the phase structure of QCD \cite{QCDphases,1,2,3,4,5,6,7,8,9,10,CFL0,CFL,m1,m1m8,Nitta,latticeQCD,diquarks,Zhitnitsky}, the behavior of quark matter in many regions of the mentioned above extended phase diagram is not known.

The topological invariant protected by CT symmetry defined in the present section is defined, strictly speaking, in the limited region of the phase diagram, where CT is not broken. Nonzero value of chemical potential breaks the charge conjugation symmetry, and therefore at $\mu \ne 0$ the invariant ${\cal N}_3^{CT}$ is not defined. However, we will see in Sect. \ref{SectGen}, that there is the topological invariant ${\cal N}_5^{\gamma^5}$ that is equal to the value of ${\cal N}_3^{CT}$ for the system of non - interacting fermions. Therefore, it represents an extension of the definition of ${\cal N}_3^{CT}$ to the general case. We expect, that there might be the place in the mentioned above extended phase diagram, where the topological phase transition to the phase with the value of   ${\cal N}_3^{CT}$ (or its extension) different from $N_c N_f=9$ is present.

To demonstrate, how this phase may be organized in principle, let us consider the  form of the Green function  of Eq. (\ref{eq:S(p)}). The terms with the transitions between the quarks are absent here. Therefore, after diagonalization we think of $A$ and $B$ as of usual functions specific for each quark. Suppose then, that for the given quark
$A(-p^2)$  has the form with $A(\infty) = 1$, $A(0) = - A_0 < 0$, and $A(P_c^2) = 0$ for a certain value of Euclidean momentum $P_c$.
We also suppose, that $B(0) <0$ while $B(P^2_c)>0$.

Let us denote $g_i(P) =A(P^2) P_i $ for $i = 1,2,3$ and $g_4(P) = B(P^2)$. The general expression for the contribution of the given quark $q$ to the topological invariant may be reduced to
\begin{eqnarray}
{\cal N}^{(q)CT}_3 &=&  N_c \frac{1}{3! \pi^2}\epsilon^{abcd} e_{\mu\nu\lambda} \,
 ~\int  d^3p
~ \tilde{g}_a \partial_{p_\mu} \tilde g_b \,
\partial_{p_\nu} \tilde g_c \, \partial_{p_\lambda}  \tilde g_d\,.\label{calMproj0}
\end{eqnarray}
where integral is taken at $\omega = 0$. Also we denoted here $\tilde{g}_a = \frac{g_a}{\sqrt{\sum_{b=1,2,3,4} g_b^2}}$.  Similar to the calculations presented in \cite{Zubkov2012,Z2016AOP} we are able to represent the last expression as
\begin{eqnarray}
{\cal N}^{(q)CT}_3 &=&  N_c \sum_l \,  {\rm sign}(g_4(P_l)) \,{\rm Res}\,(P_l)\label{calMproj}
\end{eqnarray}
Here the sum is over the positions of the common zeros of functions $g_i(P) =A(P^2) P_i $ for $i = 1,2,3$ while
\begin{eqnarray}
{\rm Res}\,(P_l) &=&  \frac{1}{8 \pi} \epsilon^{ijk}\, \int_{\Sigma(P_l)} \, \hat{g}_i  d \hat{g}_j \wedge d \hat{g}_k
\end{eqnarray}
is the integer number, the corresponding integral is  along the  infinitely small surface $\Sigma$ placed in $3D$ momentum space at $\omega = 0$, which is wrapped around the given zero of $g_i$ ($i=1,2,3$). We denoted
$\hat{g}_i = \frac{g_i}{\sqrt{\sum_{j=1,2,3}g^2_j}}$.
If some of the zeros form the two dimensional surface, this does not change the above expression. In our particular case we come to the expression
\begin{eqnarray}
{\cal N}^{(q)CT}_3 &=&    N_c {\rm sign}(B(0)) \,{\rm Res}\,(0) + N_c{\rm sign}(B(P_c)) \,{\rm Res}\,(P_c) = -N_c{\rm sign}(B(0)) + 2\,N_c {\rm sign}(B(P_c)) = 3N_c \label{calMproj2}
\end{eqnarray}
Therefore, if the functions $A$ and $B$  for $q = u,d,s$ would behave in the given way, we obtain
\begin{eqnarray}
{\cal N}^{CT}_3 &=&    3 N_f N_c \label{calMproj3}
\end{eqnarray}
Notice once again that we did not demonstrate that  the considered form of the Green function corresponds to any particular external conditions. However, such a possibility is not excluded, which means, that the topological phase transition to the state with ${\cal N}^{CT}_3 \ne N_f N_c$ may appear. It is obvious, that the interactions must be strong to provide such a transition. In Hadronic phase the interactions are strong, which suggests to search for the existence of the new topological sub - phases within this phase. The whole story may be related to the infrared physics, where strong interactions are especially strong, and therefore, the functions $A(P^2)$ and $B(P^2)$ may, possibly, become negative at $P=0$ passing through zero at the values of $P$ that are not equal to each other and may be much smaller than $\Lambda_{QCD}$.

\section{Hypothetical domain wall fermions in QCD}
\label{DomainWall}

\begin{figure}[!thb]
\begin{center}
{\includegraphics[height=60mm,width=80mm,angle=0]{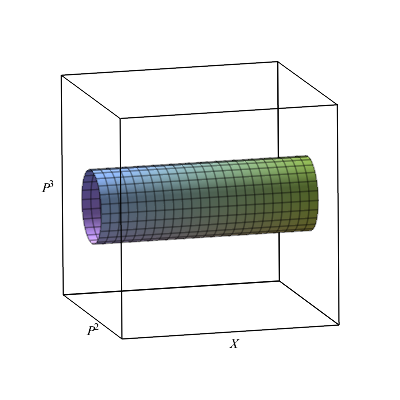}}
{\includegraphics[height=60mm,width=80mm,angle=0]{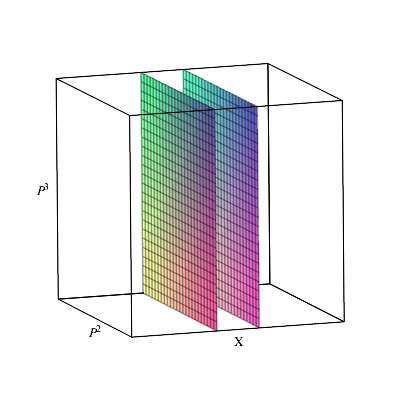}}
\end{center}
\caption{\label{fig.1}  {\bf Left:} The form of the surface  ${\cal C}\otimes R\otimes R$ in the coordinates $x,p_2,p_3$. Contour $\cal C$ embraces zero in the plane $p^2,p^3$. while the first $R$ corresponds to the values of $p$ and the second $R$ corresponds to the values of $x$. The tube crosses the domain wall placed at $x=0$. Singularity of the Green function is situated at the domain wall inside the tube. {\bf Right:} The surface ${\cal C}\otimes R\otimes R$  may be transformed to the form of the two hyperplanes ($p_1,p_2,p_3$) placed at $x = \pm \epsilon$, where $\epsilon$ is small. We represent here the slices of those hyperplanes with fixed $p_1$. The two hyperplanes are placed, therefore at the opposite sides of the domain wall.}
\end{figure}

In this section we consider the possibility that the CT symmetric phases of quark matter with different values  of ${\cal N}^{CT}_3$ coexist (one of them may be metastable). For definiteness we restrict ourselves in this section by the case of zero temperature. Notice, that the consideration of this section has the general character and it does not refer to the quark matter only. One may, actually, consider the derivation of the present section for any field theoretical or condensed matter system with the topological invariant of the form of Eq. (\ref{TopInvariantMassive}) with appropriate matrix $K$. The examples of such systems are given by the gapped phases of superfluid $^3He$, (see \cite{Volovik2003}) and by the topological insulators of various types \cite{HasanKane2010,Xiao-LiangQi2011,Schnyder2008,Kitaev2009,Grinevich1988,BevanNature1997}.

The boundary between the two phases is the domain wall. It should carry the massless fermions. Their number is equal to the jump of the topological invariant $N_{CT}$ across the interface. This may be proved as follows (for the description of the method see also \cite{Volovik2003} and \cite{EssinGurarie2011}).

Let us consider the case, when the domain wall forms the plane $yz$. Momenta $p^2,p^3$ along the plane remain the good quantum numbers, and they may be considered as parameters on which the Green function $\hat{\cal G}$ depends. Thus in this case the Green function $\hat{\cal G}$ represents the operator depending functionally on the momentum  operator $\hat{p}^1 $, and on the coordinate operator $r_1$ as well as on the parameters $p^2,p^3$, which represent conserved momenta. Below we imply $p^4=0$ and consider the following expression
\begin{equation}
{\cal N}_1({\cal C}) = \frac{1}{2\pi i}\, \int_{{\cal C}} {\rm Tr}\,K\, \hat{\cal G}(p^2, p^3) d \hat{\cal G}^{-1}(p^2,p^3)\label{N19}
\end{equation}
Here $K={\bf K}_{CT}$, contour $\cal C$ embraces zero in the plane $p^2,p^3$.  This quantity counts the number of the zero modes incident at the domain wall. This may be demonstrated as follows. At $\omega = 0$ we denote
$$\hat{\cal G} =  - \sum_n {\cal E}_n(p_2,p_3)|n,p_2,p_3\rangle\langle n,p_2,p_3|$$ where the sum is over the eigenstates of the Hamiltonian enumerated by index $n$ and the values of conserved momenta $p_2,p_3$. Let us denote ${G}_n =  - {\cal E}_n(p_2,p_3)$. Matrix $K$ commutes with $\cal G$, therefore, they are diagonalized simultaneously, and the corresponding eigenvalues of $K$ are denoted by $K_n$. Then
Eq. (\ref{N19}) may be represented as
\begin{equation}
{\cal N}_1({\cal C}) = \frac{1}{2\pi i}\, \sum_n \, \int_{{\cal C}} \, K_n {G}_n d {G}^{-1}_n \label{N129}
\end{equation}
This expression gives the number of the massless states (or zeros of the Green function) protected by $CT$ symmetry. Only those states which are localized at the domain wall  may have gapless excitations (of correspond to zeros of the Green function) because in the bulk of the given system there are no poles and zeros of the Green function. Thus we come to the conclusion, that ${\cal N}_1$ enumerates the gapless fermion modes and modes corresponding to the zeros of the Green function.

We use the Wigner transform of the Green function in order to describe the correspondence between the bulk and the domain wall. Wigner transform \cite{Wigner} of the Green function now depends on the parameters $p^2,p^3$
\begin{equation}
 \tilde{G}(R,p,p^2,p^3) = \int d{ r} e^{-i { p} r} G(R+{r}/2,{ R}-{ r}/2)\label{W}
\end{equation}
where $G({ r}_1, { r}_2)$ represents matrix elements of $\hat{\cal G}(p^2,p^3)$. Here ${ R}$ and ${ p}$  correspond to the $x$ coordinate, that is $R =x$, $p = p_1$. One may rewrite expression for $N_1$ as follows (for the details see \cite{Z2016AOP})
\begin{equation}
{\cal N}_1({\cal C}) = \frac{1}{2\pi i}\, \int d { R} \frac{d { p}}{(2\pi)} \int_{{\cal C}} {\rm Tr}\, K\,\tilde{G}({ R},{ p},p^2,p^3) d \Big[\tilde{G}^{(0)}({ R},{ p},p^2,p^3)\Big]^{-1}\label{N1119}
\end{equation}
Here we denote by $ \tilde{G}^{(0)}({ R},{ p},p^2,p^3)$ the zero order approximation to $\tilde{G}({ R},{ p},p^2,p^3) $ in the derivative expansion given by
$$
\tilde G^{(0)}({ R},{ p},p^2,p^3) =  \frac{1}{{\cal Q}({ R},{ p},p^2,p^3)}
$$
Here function $\cal Q$ represents the  Weyl symbol of operator $\hat{\cal Q} \equiv {\cal G}^{-1}$  being the Wigner transform of its matrix elements \cite{Weyl,berezin}.
 The mentioned derivative expansion gives the iterative solution for the Groenewold equation (we omit parameters $p^2,p^3$)
\begin{eqnarray}
1 &=& {\cal Q}({ R},{ p})*\tilde G({ R},{ p})\nonumber\\ && \equiv {\cal Q}({ R},{ p})e^{\frac{i}{2}(\overleftarrow{\partial}_{ R}\overrightarrow{\partial}_{ p} - \overleftarrow{\partial}_{ p}\overrightarrow{\partial}_{ R})}\tilde G({ R},{ p})  \label{id}
\end{eqnarray}
We use Eq. (\ref{id}) and expand exponent in powers of its arguments, which gives
\begin{eqnarray}
\tilde G({ R},{ p})  &= &\tilde G^{(0)}({ R},{ p}) + \tilde G^{(1)}({ R},{ p}) + \tilde G^{(2)}({ R},{ p}) + ...  \label{Gexp2}\\
\tilde G^{(1)}  &= &\frac{i}{2} \tilde G^{(0)} \frac{\partial \Big[\tilde G^{(0)}\Big]^{-1}}{\partial { p}} \tilde G^{(0)}  \frac{\partial  \Big[\tilde G^{(0)}\Big]^{-1}}{\partial { R}} \tilde G^{(0)}
\nonumber\\
 & &-\frac{i}{2} \tilde G^{(0)} \frac{\partial \Big[\tilde G^{(0)}\Big]^{-1}}{\partial { R}} \tilde G^{(0)}  \frac{\partial  \Big[\tilde G^{(0)}\Big]^{-1}}{\partial { p}} \tilde G^{(0)}
\end{eqnarray}
Next, let us substitute Eq. (\ref{Gexp2}) into Eq. (\ref{N111}). Since Eq. (\ref{N1}) is the topological number, in the whole series of terms entering our expression only the topological term contributes. The only topological term we are able to find in this derivative expansion is the following one
\begin{eqnarray}
{\cal N}_1({\cal C})&\equiv& {\cal N}_3 =   \frac{1}{3!\,(2\pi)^2}\,\,\int_{} \,{\rm Tr} \Big[K\,  \tilde{G}^{(0)} d (\tilde{G}^{(0)})^{-1}\wedge d  \tilde{G}^{(0)} \wedge d(\tilde{G}^{(0)})^{-1} \Big]  \label{nuGHall229}
\end{eqnarray}
Here the integral is over the surface ${\cal C}\otimes R\otimes R$, where the first $R$ corresponds to the values of $p$ while the second $R$ corresponds to the values of $x$.

Since there are no poles of $\tilde{G}$ inside the bulk we are able to deform this hyper - surface into the form of the
 closed surface, which embraces the line $x =0 $, $p_1=p_2 = p_3= 0$. Such a surface may be chosen in the form of the two planes placed at $x = \pm \epsilon$, where $\epsilon$ is small. This transformation is illustrated by Fig. \ref{fig.1}. This gives the difference between the values of ${ N}_{CT}$ at the two sides of the domain wall.

\section{Color Flavor Locking (CFL) phase}
\label{SectCFL}

It is widely believed that the ground state of the quark - gluon matter at extremely large baryon chemical potential is  the color - flavor locking phase (CFL).  In the simplest phenomenological models of this phase the three quarks $u,d,s$ are supposed to be massless. The condensate is formed  \cite{CFL0, CFL}
\begin{equation}
\langle [\psi^{i}_{\alpha}]^t i \gamma^2\gamma^0 \gamma^5 \psi^j_{\beta} \rangle \sim \Phi^I_J \, \epsilon_{\alpha \beta J} \epsilon^{ijI} \sim  \epsilon_{\alpha \beta I} \epsilon^{ijI}\label{CFL}
\end{equation}
The condensate of Eq. (\ref{CFL}) appears in \cite{CFL0} in the form of $\langle [\psi^{i}_{\alpha}]^t C \gamma^5 \psi^j_{\beta} \rangle$, for the definition of charge conjugation matrix $C$ the authors of \cite{CFL0} refer to \cite{spinors}, where in Eqs. (1.27), (1.33) it is defined as $C = - i \gamma^0 \gamma^2$.

We are not aware of the exhaustive proof that this phase indeed appears at large baryon chemical potential. The most popular explanations of the appearance of the CFL phase at asymptotically large values of baryon chemical potential $\mu_B$ are based on the consideration of the one - gluon exchange \cite{Schafer}.

 However, even at very large values of  $\mu_B$ the strong interaction might remain strong.  Strictly speaking, we are not aware of neither the proof that at the large chemical potential the nonperturbative effects disappear, nor of the proof that they remain. Therefore, we may rely on the known results at vanishing chemical potential and large temperature. Naively, one would expect, that in the running coupling constant $\alpha_s$ the scale equals to temperature is to be substituted. If it is sufficiently large, then the coupling constant is sufficiently small. This indicates that the nonperturbative effects may disappear. However, numerical lattice simulations show that this is incorrect. Even at large temperatures the spacial string tension survives \cite{sigma_space}, which is a purely nonperturbative effect.
At large chemical potential and vanishing temperature one should substitute $\mu_B$ to the running coupling constant $\alpha_s$, the value of $\alpha_s(\mu_B)$ is relatively small compared to the value of $\alpha(\Lambda_{QCD})$, but this does not mean that the nonperturbative effects disappear. Although the QCD at zero temperature and large $\mu_B$ is understood much weaker, one might suppose, that here the spatial string tension survives as well. This means that the nonperturbative effects still may be important. Unfortunately, this region of the phase diagram is not accessible for the lattice numerical simulations, which complicates the situation. Nevertheless, following the authors of \cite{CFL0,CFL,m1,m1m8,Nitta} we assume, that the CFL phase appears at least at the asymptotically large values of baryonic chemical potential.

There are $18$ scalar fluctuations of $\Phi$ around this condensate (there are also $18$ pseudo - scalar fluctuations with the same masses \cite{diquarks}). The symmetry breaking pattern is $SU(3)_L \otimes SU(3)_R \otimes SU(3)_C\otimes U(1)_V \rightarrow SU(3)_{CF}/Z_3$. That's why there are $8+8+1$ massless Goldstone modes.  Among the remaining $9+9+1$ Higgs modes there are two octets of the traceless modes and two singlet  trace modes as well as one state that becomes massive due to the instantons.
Correspondingly, the quark excitations also form singlets and octets. The singlet fermionic gap $\Delta_{1}$ is twice larger than the octet fermionic gap $\Delta_{8}$ (see Sect. 5.1.2. of \cite{CFL}).

In terms of the left - handed $q_L$ and the right - handed $q_R$ components of quarks one may represent Eq. (\ref{CFL}) using chiral representation of Dirac matrices as follows
\begin{equation}
\langle q^{Ai}_{L,\alpha} \epsilon_{AB}  q^{Bj}_{L,\beta} \rangle =-\langle \bar{q}^{Ai}_{L,\alpha} \epsilon_{AB}  \bar{q}^{Bj}_{L,\beta} \rangle =  \langle q^{Ai}_{R,\alpha} \epsilon_{AB}  q^{Bj}_{R,\beta} \rangle  = -  \langle \bar{q}^{Ai}_{R,\alpha} \epsilon_{AB}  \bar{q}^{Bj}_{R,\beta} \rangle \sim \delta^I_J \, \epsilon_{\alpha \beta J} \epsilon^{ijI}\label{CFL2}
\end{equation}
Spinor notations and relative signs in those expressions correspond to those of Eq. (2.1) in \cite{CFL0}, where $\epsilon^{ab} = \left(\begin{array}{cc}0 & 1\\-1 & 0 \end{array} \right)$ while $\epsilon_{\dot{a}\dot{b}} = \left(\begin{array}{cc}0 & -1\\1 & 0  \end{array}\right)$ according to the notations of Eqs. (1.34), (1.35) in  \cite{spinors}, to which the authors of \cite{CFL0} refer for the explanation of their own notations. It is worth mentioning, that $\Big[ q^{Ai}_{L,\alpha} \epsilon_{AB}  q^{Bj}_{L,\beta} \Big]^+ =-\Big[ \bar{q}^{Bj}_{L,\beta} \epsilon_{BA} \bar{q}^{Ai}_{L,\alpha} \Big]$ (and the similar expression for the operators of the right handed fermions) by definition of the hermitian conjugation.

Instead of $q_R,q_L$ one may use the CP - conjugated spinors $q_R^{c} = i \sigma^2 \bar{q}_R$, $q_L^{c} = -i \sigma^2 \bar{q}_L$. In terms of these spinors we get
\begin{equation}
-\langle q^{c,Ai}_{L,\alpha} \epsilon_{AB}  q^{c,Bj}_{L,\beta} \rangle =- \langle q^{c,Ai}_{R,\alpha} \epsilon_{AB}  q^{c,Bj}_{R,\beta} \rangle   \sim \delta^I_J \, \epsilon_{\alpha \beta J} \epsilon^{ijI}\label{CFL02}
\end{equation}

We use the conventional definition of the corresponding Nambu - Gorkov spinors:  ${\cal Q}_L = (-q^{c}_L, q_L)^T = (i \sigma^2 \bar{q}_L, q_L)^T$, ${\cal Q}_R = (q_R, q^{c}_R)^T= (q_R, i \sigma^2 \bar{q}_R)^T$. The condensate may be written as
\begin{equation}
-\langle \bar{Q}^{c,i}_{L,\alpha} Q^{c,j}_{L,\beta} \rangle =- \langle \bar{Q}^{c,i}_{R,\alpha}  Q^{c,j}_{R,\beta} \rangle   \sim \delta^I_J \, \epsilon_{\alpha \beta J} \epsilon^{ijI}\label{CFL12}
\end{equation}
where
\begin{equation}
\bar{\cal Q}_L = {\cal Q}_L^Ti \gamma^2\gamma^0, \quad \bar{\cal Q}_R = -{\cal Q}_R^Ti \gamma^2\gamma^0
\end{equation}
In the simplest Nambu - Jona - Lasinio model of CFL phase \cite{CFL} the effective interaction 4 - fermion lagrangian has the form
\begin{eqnarray}
&&L_{qq} = -{\rm const}\, \sum_{A,A^\prime = 2,5,7} \Big(\bar{\psi}^{i}_{\alpha} i \gamma^2\gamma^0 \gamma^5 \hat\lambda_C^A \hat \lambda_F^{A^\prime} [\bar{\psi}^T]^j_{\beta}\Big)\Big([\psi^{i}_{\alpha}]^T i \gamma^2\gamma^0 \gamma^5 \hat \lambda_C^A\hat \lambda_F^{A^\prime} \psi^j_{\beta}\Big)\label{F4}\\ &&
= {\rm const}\, \sum_{A,A^\prime = 2,5,7} \Big([\bar{q}^{i}_{L,\alpha}]^T i \sigma^2 \hat \lambda_C^A\hat \lambda_F^{A^\prime} \bar{q}^j_{L,\beta}+[\bar{q}^{i}_{R,\alpha}]^T i \sigma^2 \hat \lambda_C^A\hat \lambda_F^{A^\prime} \bar{q}^j_{R,\beta}\Big)\Big([q^{i}_{L,\alpha}]^T i \sigma^2 \hat \lambda_C^A\hat \lambda_F^{A^\prime} q^j_{L,\beta}+[q^{i}_{R,\alpha}]^T i \sigma^2 \hat \lambda_C^A\hat \lambda_F^{A^\prime} q^j_{R,\beta}\Big)\nonumber
\end{eqnarray}
where $\hat \lambda_C^A$ and $\hat \lambda_F^A$ are the  generators of $SU(3)_C$ and $SU(3)_F$ respectively. Here $i \hat \lambda^2_{ij} = \epsilon_{ij3}$, $i \hat \lambda^5_{ij} = \epsilon_{i2j}$, $i \hat \lambda^7_{ij} = \epsilon_{1ij}$.   There is one more form of the CFL condensate: $$ -\langle[q^{i}_{L,\alpha}]^T i \sigma^2 \hat \lambda_C^A\hat \lambda_F^{A^\prime} q^j_{L,\beta}+[q^{i}_{R,\alpha}]^T i \sigma^2 \hat \lambda_C^A\hat \lambda_F^{A^\prime} q^j_{R,\beta}\rangle \sim  \delta^{A A^\prime}, \quad A,A^\prime = 2,5,7$$
In the presence of this condensate the mean field lagrangian for the left - handed fermionic excitations may be written as follows:
\begin{equation}
L_L =\frac{1}{2} \bar{\cal Q}^i_{L,\alpha}\Big( (\gamma^\mu p_\mu - \mu \gamma^0 \gamma^5)\delta_{ij}\delta^{\alpha \beta}  - \epsilon_{ijI}\epsilon^{\alpha\beta I}M\Big) {\cal Q}^j_{L,\beta}\label{leftq}
\end{equation}
Here $M$ is the parameter of the dimension of mass, that is proportional to the condensate (coefficient of proportionality is the constant entering the four fermion interaction term of Eq. (\ref{F4})). $\mu$ is the quark chemical potential.  In the basis of the Nambu - Gorkov spinors the time reversal transformation $q_{L,R}(t) \to i \sigma^2 \bar{q}_{L,R}(-t)$ corresponds to matrix
\begin{equation}
{\cal K}_T = - \gamma^0\gamma^5\label{KT0}
\end{equation}
In the following we will use the operator that differs from Eq. (\ref{KT0}) by factor   $i$:
\begin{equation}
{\bf K}_T = i {\cal K}_T = - i \gamma^0\gamma^5\label{KT1}
\end{equation}
It also commutes with the Green function at $\omega= 0$ and therefore may be used to compose the topological invariants.

The similar situation takes place for the right - handed quarks. In terms of the Nambu - Gorkov spinors the mean field lagrangian for the massive right - handed quarks  may be written as:
\begin{equation}
L_R =\frac{1}{2}\bar{\cal Q}^i_{R,\alpha}\Big( (\gamma^\mu p_\mu + \mu \gamma^0 \gamma^5)\delta_{ij}\delta^{\alpha \beta}  - \epsilon_{ijI}\epsilon^{\alpha\beta I}M\Big) {\cal Q}^j_{R,\beta}\label{rightq}
\end{equation}
Therefore, the matrix of time reversal transformation is given by
\begin{equation}
{\bf K}_T = i \gamma^0\gamma^5\Gamma^5
\end{equation}
where $\Gamma^5 = +1$ for the right - handed spinors and $\Gamma^5 = -1$ for the left - handed.

Adding the Dirac mass matrix $m$ we arrive at the tree level lagrangian
\begin{equation}
L =\frac{1}{2}\bar{\cal Q}^i_{\alpha}\Big( (\gamma^\mu p_\mu + \mu \gamma^0 \gamma^5\Gamma^5 - m \gamma^5 \Gamma^0 \Gamma^5)\delta_{ij}\delta^{\alpha \beta}  - \epsilon_{ijI}\epsilon^{\alpha\beta I}M\Big) {\cal Q}^j_{\beta}\label{leftrightq}
\end{equation}
If $m$ would be the same for all fermions, the dispersion of quasiparticles described by this lagrangian is given by
$$
{\cal E} = \pm \sqrt{M^2 + \Big(\mu \pm \sqrt{\vec{p}^2 + m^2}\Big)^2}, \, \pm \sqrt{4 M^2 + \Big(\mu \pm \sqrt{\vec{p}^2 + m^2}\Big)^2}
$$
One can see, that for $M = 0, \mu < m$ the dispersion matches qualitatively the situation that takes place in hadronic phase, where the fermions are gapped, and the Dirac mass is present. For $M=0, \mu > m$ the Fermi surface is formed. For $M\ne 0$ the system is gapped both for $\mu < m$ and $m < \mu$. This case qualitatively corresponds to the CFL phase.

The lagrangian of Eq. (\ref{leftrightq}) appears if we omit all interactions and take into account only the contribution of the quark - quark condensate to the Majorana mass term of quarks. The real effective lagrangian is much more complicated than Eq. (\ref{leftrightq}), it contains extra terms due to interactions.
However, we may assume, that the system may be deformed continuously to the one having the complete lagrangian of the form of  Eq. (\ref{leftrightq}).

We may consider in the CFL phase the topological invariant protected by T symmetry. (Notice, that in the basis of the considered here Nambu - Gorkov spinors the charge conjugation symmetry acts as follows: $Q \to \gamma^5 \Gamma^0 Q$,  the composition of C and T acts as $
Q \to {\bf K}_{CT} Q = - i \gamma^0 \Gamma^0 \Gamma^5 Q
$ while the P transformation is given by matrix $\gamma^0 \Gamma^0$. Here $\Gamma^0 = \left(\begin{array}{cc}0 & 1\\1 & 0 \end{array} \right)$.)
Let us calculate the corresponding topological invariant
\begin{equation}
{\cal N}^{T}_3 = {e_{ijk}\over{48\pi^2}} ~
{\bf tr}\left[  \int_{\omega=0}   d^3p ~{\bf K}_{T}
~G\partial_{p_i} G^{-1}
G\partial_{p_j} G^{-1} G\partial_{p_k} G^{-1}\right]\,.
\label{3DTopInvariant_tau_24e28}
\end{equation}
For the Green function of the form
\begin{equation}
{G}^{-1} =  \gamma^\mu p_\mu - m \gamma^5 \Gamma^0 \Gamma^5 +  \mu \gamma^0 \gamma^5 \Gamma^5  - {\bf E} M
\end{equation}
with
${\bf E}_{ij}^{\alpha\beta} = \epsilon_{ijI}\epsilon^{\alpha\beta I}$
we easily obtain
$$
{\cal N}^{T}_3  = 0
$$
One can also easily check that for the noninteracting fermions with Dirac mass the topological invariant protected by the time reversal symmetry vanishes as well. Further, in Sect. \ref{SectGen} we will introduce the topological invariant that takes nontrivial values in both cases.

It is worth mentioning, that the topological properties of the CFL phase have been considered in \cite{Nishida:2010wr}, where bulk topological invariant (Eqs. (15) and (16) of \cite{Nishida:2010wr}) has been considered for the case when the system may be deformed continuously to the noninteracting one (that topological invariant, though, is not expressed directly through the Green function).

\section{Relation between the fermion zero modes that reside on vortices and the topological invariant in mixed $r-p$ space}
\label{SectVORT}


\begin{figure}[!thb]
\begin{center}
{\includegraphics[height=60mm,width=80mm,angle=0]{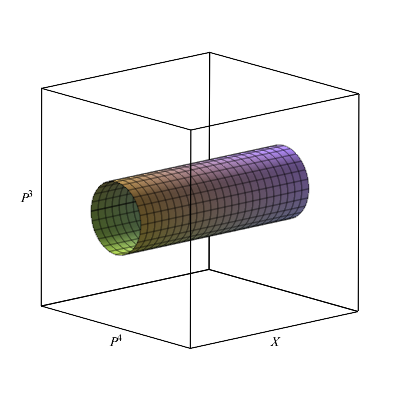}}
{\includegraphics[height=60mm,width=80mm,angle=0]{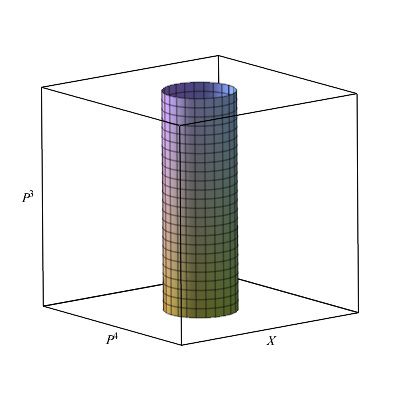}}
\end{center}
\caption{\label{fig.2}  {\bf Left:} The form of the surface  ${\cal C}\otimes R^2\otimes R^2$ in the coordinates $x,p_3,p_4$. Contour $\cal C$ embraces zero in the plane $p^3,p^4$. while the first $R^2$ corresponds to the values of $x,y$ and the second $R^2$ corresponds to the values of $p_1,p_2$.  Singularity of the Green function is situated at the vortex ($x=y=0$) inside the tube. {\bf Right:} The tube ${\cal C}\otimes R^2\otimes R^2$  may be  rotated first such that the circle $\cal C$ is transformed to the circle in the plane $x,p_4$. This is visualized here as the result of the rotation around the axis $p_4$.}
\end{figure}

To reveal the relation of momentum space topological invariants with the zero modes of the fermions that reside on vortices we develop the method of Chapter 23 of \cite{Volovik2003}. In this section we consider the system of general type, not necessarily the one of the QCD in the CFL phase. Let us consider the case, when the vortex forms the straight line directed along the $z$ axis. Momentum $p^3$ along the vortex remains the good quantum number, and it may be considered as parameter on which the Green function $\hat{\cal G}$ depends. The same is valid in the static case for the imaginary frequency $p^4$. Thus in this case the Green function $\hat{\cal G}$ represents the operator depending functionally on the momentum  operators $\hat{p}^1,\hat{p}^2$, and on the coordinate operators $r_1$, $r_2$ as well as on the parameters $p^3,p^4$, which represent conserved momentum and imaginary frequency. We then consider the following expression
\begin{equation}
{\cal N}_1({\cal C}) = \frac{1}{2\pi i}\, \int_{{\cal C}} {\rm Tr}\, \hat{\cal G}(p^3, p^4) d \hat{\cal G}^{-1}(p^3,p^4)\label{N1}
\end{equation}
Here contour $\cal C$ embraces zero in the plane $p^3,p^4$.  This quantity encodes the chirality of the zero modes incident at the vortex core. This may be demonstrated as follows. For simplicity we may consider here the case of the noninteracting system. The generalization to the interacting system is straightforward.
Now we have $\hat{\cal G} = i\omega - \hat{H}$, where $\omega = p^4$ is imaginary frequency, and
$$\hat{\cal G} = i\omega - \sum_n {\cal E}_n(p_3)|n,p_3\rangle\langle n,p_3|$$ where the sum is over the eigenstates of the Hamiltonian enumerated by index $n$ and the values of conserved momentum $p_3$. Let us denote ${G}_n = i\omega - {\cal E}_n(p_3)$. Then
Eq. (\ref{N1}) may be represented as
\begin{equation}
{\cal N}_1({\cal C}) = \frac{1}{2\pi i}\, \sum_n \, \int_{{\cal C}} \, {G}_n d {G}^{-1}_n \label{N12}
\end{equation}
This expression gives the sum of the chiralities of the states enumerated by $n$. Only those states which are localized in the vortex core may have gapless excitations because in the bulk of the given system there are no poles of the Green function. Thus we come to the conclusion, that ${\cal N}_1$ enumerates the gapless fermion modes residing at the vortex core, and its sign corresponds to their chirality.

Let us use the Wigner transform of the Green function in order to describe the correspondence between the bulk and the vortex sheet (for the description of the method see Chapter 23 in \cite{Volovik2003}).
For the description of the properties of the Wigner transformation of Green functions see \cite{Z2016,Z2016AOP} and references therein. Wigner transform \cite{Wigner} of the Green function is defined as
\begin{equation}
 \tilde{G}({\bf R},{\bf p},p^3,p^4) = \int d^D{\bf r} e^{-i {\bf p} {\bf r}} G({\bf R}+{\bf r}/2,{\bf R}-{\bf r}/2)\label{W}
\end{equation}
where $G({\bf r}_1, {\bf r}_2)$ represents matrix elements of $\hat{\cal G}(p^3,p^4)$. Here ${\bf R}$ and ${\bf p}$ are two dimensional and correspond to $x,y$ coordinates, that is ${\bf R} = (x,y)$ and ${\bf p} = (p_1,p_2)$ . One may rewrite expression for ${\cal N}_1$ as follows
\begin{equation}
{\cal N}_1({\cal C}) = \frac{1}{2\pi i}\, \int d^2 {\bf R} \frac{d^2 {\bf p}}{(2\pi)^2} \int_{{\cal C}} {\rm Tr}\, \tilde{G}({\bf R},{\bf p},p^3,p^4) d \Big[\tilde{G}^{(0)}({\bf R},{\bf p},p^3,p^4)\Big]^{-1}\label{N111}
\end{equation}
Here we denote by $ \tilde{G}^{(0)}({\bf R},{\bf p},p^3,p^4)$ the zero order approximation to $\tilde{G}({\bf R},{\bf p},p^3,p^4) $ in the derivative expansion. The latter expansion gives the iterative solution for the Groenewold equation (we omit parameters $p^4,p^3$)
\begin{eqnarray}
1 &=& {\cal Q}({\bf R},{\bf p})*\tilde G({\bf R},{\bf p})\nonumber\\ && \equiv {\cal Q}({\bf R},{\bf p})e^{\frac{i}{2}(\overleftarrow{\partial}_{\bf R}\overrightarrow{\partial}_{\bf p} - \overleftarrow{\partial}_{\bf p}\overrightarrow{\partial}_{\bf R})}\tilde G({\bf R},{\bf p})  \label{id}
\end{eqnarray}
Here function $\cal Q$ represents the so - called Weyl symbol of operator $\hat{\cal Q} \equiv {\cal G}^{-1}$  being the Wigner transform of its matrix elements \cite{Weyl,berezin}.
We use Eq. (\ref{id}) and expand exponent in powers of its arguments, which gives
\begin{eqnarray}
\tilde G({\bf R},{\bf p})  &= &\tilde G^{(0)}({\bf R},{\bf p}) + \tilde G^{(1)}({\bf R},{\bf p}) + \tilde G^{(2)}({\bf R},{\bf p}) + ...  \label{Gexp2}\\
\tilde G^{(1)}  &= &\frac{i}{2} \tilde G^{(0)} \frac{\partial \Big[\tilde G^{(0)}\Big]^{-1}}{\partial {\bf p}} \tilde G^{(0)}  \frac{\partial  \Big[\tilde G^{(0)}\Big]^{-1}}{\partial {\bf R}} \tilde G^{(0)}
\nonumber\\
 & &-\frac{i}{2} \tilde G^{(0)} \frac{\partial \Big[\tilde G^{(0)}\Big]^{-1}}{\partial {\bf R}} \tilde G^{(0)}  \frac{\partial  \Big[\tilde G^{(0)}\Big]^{-1}}{\partial {\bf p}} \tilde G^{(0)}
\nonumber\\
\tilde G^{(2)}  &= &\frac{i}{2} \tilde G^{(0)} \frac{\partial \Big[\tilde G^{(0)}\Big]^{-1}}{\partial {\bf p}}  \frac{\partial  \Big[\tilde G^{(1)}\Big]}{\partial {\bf R}} \nonumber\\&&
-\frac{i}{2} \tilde G^{(0)} \frac{\partial \Big[\tilde G^{(0)}\Big]^{-1}}{\partial {\bf R}}  \frac{\partial  \Big[\tilde G^{(1)}\Big]}{\partial {\bf p}} \nonumber\\&&
-\frac{1}{2}\Big(\frac{i}{2} \Big)^2 \tilde G^{(0)}\Big( \frac{\partial}{\partial R_i} \frac{\partial}{\partial {R_j}} \Big[\tilde G^{(0)}\Big]^{-1} \Big)\Big( \frac{\partial}{\partial p_i}\frac{\partial}{\partial p_j}\tilde G^{(0)}\Big)
 \nonumber\\&&
-\frac{1}{2}\Big(\frac{i}{2} \Big)^2 \tilde G^{(0)}\Big( \frac{\partial}{\partial p_i} \frac{\partial}{\partial {p_j}} \Big[\tilde G^{(0)}\Big]^{-1} \Big)\Big( \frac{\partial}{\partial R_i}\frac{\partial}{\partial R_j}\tilde G^{(0)}\Big)
 \nonumber\\&&
+\Big(\frac{i}{2} \Big)^2 \tilde G^{(0)}\Big( \frac{\partial}{\partial R_i} \frac{\partial}{\partial {p_j}} \Big[\tilde G^{(0)}\Big]^{-1} \Big)\Big( \frac{\partial}{\partial p_i}\frac{\partial}{\partial R_j}\tilde G^{(0)}\Big)
\end{eqnarray}
Here $\tilde G^{(0)}({\bf R},{\bf p})$ is given by the solution of equation
\begin{eqnarray}
&& \tilde G^{(0)}({\bf R},{\bf p}) {\cal Q}({\bf R}, {\bf p}) =1 \label{Q0}
\end{eqnarray}
Next, let us substitute Eq. (\ref{Gexp2}) into Eq. (\ref{N111}). Since Eq. (\ref{N1}) is the topological number, in the whole series of terms entering our expression only the topological term contributes. The only topological term we are able to find in our expression is the following one
\begin{eqnarray}
{\cal N}_1({\cal C})&\equiv& {\cal N}_5 =   \frac{2}{5!\,(2\pi)^3}\,\,\int_{} \,{\rm Tr} \Big[  \tilde{G}^{(0)} d (\tilde{G}^{(0)})^{-1}\wedge d  \tilde{G}^{(0)} \wedge d(\tilde{G}^{(0)})^{-1} \wedge d  \tilde{G}^{(0)} \wedge d(\tilde{G}^{(0)})^{-1}\Big]  \label{nuGHall22}
\end{eqnarray}
Here the integral is over the surface ${\cal C}\otimes R^2 \otimes R^2$, where the first $R^2$ corresponds to the values of $p_1,p_2$ while the second $R^2$ corresponds to the values of $x,y$.  The topological invariant of the form of Eq. (\ref{nuGHall22}) has been discussed earlier for the particular case of vortices in $^3$He-B in \cite{Silaev}, where instead of our well defined (through the Wigner transformation) $G^{(0)}$ the Green function depending both on coordinates and on momenta was introduced intuitively. Actually our construction may be considered as the precise formulation of the one discussed in \cite{Silaev}.

\begin{figure}[!thb]
\begin{center}
{\includegraphics[height=60mm,width=80mm,angle=0]{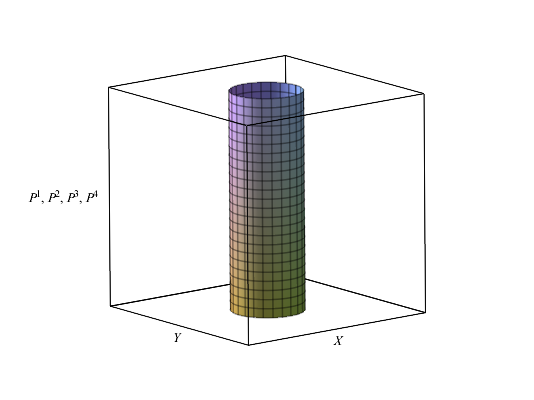}}
\end{center}
\caption{\label{fig.3} The final form of the integration surface for Eq. (\ref{nuGHall22}): ${\cal C}\otimes R^4$, where $\cal C$ is the circle in the plane $x, y$ that embraces the position of the vortex. $R^4$ is the 4 - dimensional momentum space.  }
\end{figure}

Since there are no poles of $\tilde{G}$ inside the bulk we are able to deform this hyper - surface into the form of the
 closed surface, which embraces the line $x = y =0 $, ${\bf p} = 0$. Such a surface may be chosen in the form of the product $S^1 \otimes R^1 \otimes R^3$, where circle $S^1$ is in the $xy$ plane with the center at $x=y=0$, $R^1$ is the line of imaginary frequency while $R^3$ is the three - dimensional momentum space
 $(p_1,p_2,p_3)$. Geometrically this transformation of the surface looks like the rotation of the tube ${\cal C}\otimes R^2 \otimes R^2$ in such a way, that the circle ${\cal C}$ in the plane $p_3,p_4$ is transformed to the circle in the plane $x, y$. This rotation may be visualized as a sequence of two steps. First, the tube is rotated in the plane $(x,p_3)$. This step is illustrated by Fig. \ref{fig.2}. As a result the circle $\cal C$ is rotated to the plane $(p_4,x)$. Next, the tube is rotated in the plane $(y,p_4)$. As a result $\cal C$ belongs to the plane $(x,y)$ - see Fig. \ref{fig.3}. Both rotations are performed in such a way, that the singularity of the Green function placed at the vortex $x=y=0$ remains inside the tube.

  Thus we came to conclusion, that the chirality of the fermion zero modes residing on the vortex (given by Eq. (\ref{N1})) is equal to the topological invariant in mixed $r-p$ space of Eq. (\ref{nuGHall22}), where the integration surface is ${\cal C}\otimes R^4$, where $\cal C$ is the circle in the plane $x, y$ that embraces the position of the vortex, while $R^4$ is the 4 - dimensional momentum space.

\section{Fermion zero modes on the vortices in the CFL phase}
\label{SectCFLVORT}


In the CFL phase the symmetry breaking pattern is $SU(3)_L \otimes SU(3)_R \otimes SU(3)_C\otimes U(1)_V \rightarrow SU(3)_{CF}/Z_3$. The first homotopic group of the factor space ${\cal H} = SU(3)_L \otimes SU(3)_R \otimes U(1)_V\otimes Z_3$ is nontrivial due to the $U(1)$ constituent. This allows the existence of topologically stable vortices.
  Also there exist the ones, which are topologically stable only in the limit, when the contribution of instantons is neglected. The latter is believed to be suppressed at high densities \cite{4} as we mentioned above. These vortexes correspond to the approximate $U(1)_A$. Below we consider first of all these approximately stable vortices. Let us denote
  \begin{equation}
\Phi^I_{L,J}= \, \epsilon_{\alpha \beta J} \epsilon^{ijI}  \langle q^{Ai}_{L,\alpha} \epsilon_{AB}  q^{Bj}_{L,\beta} \rangle , \quad \Phi^I_{R,J}= - \, \epsilon_{\alpha \beta J} \epsilon^{ijI}  \langle q^{Ai}_{R,\alpha} \epsilon_{AB}  q^{Bj}_{R,\beta}\rangle \label{CFL8}
\end{equation}
The simplest configuration of the $U(1)_B$  vortex corresponds to  \cite{Nitta}
 $$\Phi_L(x,y,z) = - \Phi_R(x,y,z)  = e^{i \theta n} f(\rho) M$$
where $\theta$ and $\rho$ are the polar coordinates in the $xy$ plane. Function $f(\rho)$ tends to unity at $\rho \rightarrow \infty$.
Similarly the simplest configuration of the $U(1)_A$  vortex is \cite{Nitta}
 \begin{equation}\bar{\Phi}_L(x,y,z) = - {\Phi}_R(x,y,z)  = e^{i \theta n} f(\rho) M \label{U1A}\end{equation}
  We suppose, that the system in the presence of the order parameter field may be continuously deformed to the one with the tree level propagator (since we are considering the CFL phase, where $\alpha_S(\mu_B)$ is essentially smaller, than the typical values of $\alpha_S$ in Hadronic phase, it is generally implied, that this continuous deformation may indeed be performed). Then the mass term for the quarks in the presence of the  condensate of the form of Eq. (\ref{U1A}) is
   \begin{eqnarray}
{\cal M} &=& e^{-i \theta n} f(\rho) M \, \epsilon_{\alpha \beta I} \epsilon^{ijI}   q^{Ai}_{L,\alpha} \epsilon_{AB}  q^{Bj}_{L,\beta}  - e^{ i \theta n} f(\rho) M     \, \epsilon_{\alpha \beta I} \epsilon^{ijI} \bar{q}^{Ai}_{L,\alpha} \epsilon_{AB} \bar{q}^{Bj}_{L,\beta} \nonumber\\&&+ e^{i \theta n} f(\rho) M \, \epsilon_{\alpha \beta I} \epsilon^{ijI}   q^{Ai}_{R,\alpha} \epsilon_{AB}  q^{Bj}_{R,\beta}  - e^{- i \theta n} f(\rho) M  \, \epsilon_{\alpha \beta I} \epsilon^{ijI}    \bar{q}^{Ai}_{R,\alpha} \epsilon_{AB} \bar{q}^{Bj}_{R,\beta}\nonumber\\
&=&  - f(\rho) M \, \epsilon_{\alpha \beta I} \epsilon^{ijI}   \bar{Q}^{i}_{L,\alpha}e^{i \gamma^5 \theta n}  Q^{j}_{L,\beta}    - f(\rho) M \, \epsilon_{\alpha \beta I} \epsilon^{ijI}   \bar{Q}^{i}_{R,\alpha}e^{i \gamma^5 \theta n}  Q^{j}_{R,\beta}
\end{eqnarray}
  Here $\gamma^5 = {\rm diag}(1,1,-1,-1)$. Then the Green function in the background of the Abelian $U(1)_A$ vortex is
\begin{equation}
\Big[ \tilde{G}^{(0)}\Big]^{-1} =  \gamma^\mu p_\mu - \mu \gamma^0 \gamma^5 \Gamma^5  - {\bf E}e^{i \gamma^5 \theta n} f(\rho) M = e^{i\theta\gamma^5n/2}\Big( \gamma^\mu p_\mu  - \mu \gamma^0 \gamma^5\Gamma^5 - {\bf E} f(\rho) M\Big)e^{i\theta \gamma^5n/2}
\end{equation}
Here
$${\bf E}_{ij}^{\alpha\beta} = \epsilon_{ijI}\epsilon^{\alpha\beta I}$$
We substitute  this expression to Eq. (\ref{nuGHall22}) and denote $\tilde{M}=f(\rho)\,M$. (Notice that instead of $p_0$ we should substitute into the Green function $i p_4$, where $p_4$ is the imaginary frequency.)
Recall, that the energy gap does not disappear if we smoothly turn off  the chemical potential. Therefore, such a deformation of theory does not lead to phase transition. During the decreasing of $\mu$ the value of any topological invariant is not changed, and as in Sect. \ref{SectCFL} we may simply set $\mu = 0$ in order to calculate ${\cal N}_1({\cal C})$ of Eq. (\ref{nuGHall22}).
The integral over $\theta$ gives the factor $2\pi n$ while the integral over the imaginary frequency gives
$$ 2 \int_0^\infty  \frac{d \omega}{(\vec{p}^2+\omega^2+{\bf E}^2\tilde{M}^2)^3}{\bf E}^2\tilde{M}^2 =\frac{3}{8\pi} \frac{{\bf E}^2\tilde{M}^2}{(\vec{p}^2+{\bf E}^2\tilde{M}^2)^{5/2}}$$
We come to
\begin{eqnarray}
{\cal N}_1({\cal C}) =  - \frac{2\times 5\times 4\times 2\pi n }{5!\,(2\pi)^3}\,\,\int_{} \,{\rm Tr} \Big[{\bf E} i\gamma^0 \gamma^5  \tilde{G}_0^{(0)} d (\tilde{G}_0^{(0)})^{-1}\wedge d  \tilde{G}_0^{(0)} \wedge d(\tilde{G}_0^{(0)})^{-1}\Big]\times \frac{\tilde{M}}{(\vec{p}^2+{\bf E}^2\tilde{M}^2)^{1/2}} \times \frac{3}{8\pi} \label{nuGHall2299}
\end{eqnarray}
Here the integral is over $3D$ momentum space while
\begin{equation}
 \tilde{G}_0^{(0)} =   \tilde{G}^{(0)}\Big|_{\theta=0}=\gamma^\mu p_\mu  - {\bf E} f(\rho) M
\end{equation}
For the considered case of the noninteracting fermions with the propagator $\tilde{G}_0^{(0)}$ we have
\begin{eqnarray}
{\cal N}_1({\cal C})&=&   2  n N_f N_c
\end{eqnarray}
One can see, that the Abelian vortex with the topological number $n$ carries $ 2 n N_f N_c$  chiral fermions (this answer does not depend on the signs of the Majorana masses of the bulk fermions).  Those fermions are Majorana because each of them corresponds to the single branch of spectrum. That means that each such fermion in the $1+1$ D  space of vortex sheet contains only one independent component and is described by one rather than two independent pair of Grassmann variables.

One may also consider the following non - Abelian vortices (it is similar to M1 in the terminology of \cite{Nitta}) with the order parameter field of the form
$$\Phi^+_L(x,y,z) =-\Phi_R(x,y,z) =  e^{i \theta n/3} e^{-i  \lambda^8 \theta n} f(\rho) M $$
where \begin{equation}
 {\lambda}^8  = \frac{1}{3}\left(\begin{array}{ccc} 1 & 0 & 0 \\0 & 1 & 0 \\ 0 & 0 & -2\end{array}\right)\label{C8}
\end{equation}
(In \cite{Nitta} the M1 vortex corresponds to $\Phi_L(x,y,z) =-\Phi_R(x,y,z) =  e^{i \theta n/3} e^{-i  \lambda^8 \theta n} f(\rho) M $.)
Now
\begin{equation}
 \tilde{G}^{-1} =  e^{i\theta\gamma^5n(1/6+\lambda_c^8/2+\lambda^8_f/2)}\Big( \gamma^\mu p_\mu -\mu \gamma^0 \gamma^5\Gamma^5 - {\bf E} f(\rho) M\Big)e^{i\theta \gamma^5n(1/6+\lambda_c^8/2 + \lambda_f^8/2)}
\end{equation}
In this situation the terms proportional to $\lambda^8$ vanish in the expression for the topological invariant and when all masses are positive, we come to
\begin{eqnarray}
{\cal N}_1({\cal C})&=&     2 n N_f  \label{nuGHall2309}
\end{eqnarray}
which means, that the number of the corresponding zero modes is equal to $2 n N_f$.

Notice, that the fermion zero modes obtained here for the mentioned two types of vortices were discussed earlier (see, for example, \cite{Nitta}) using the direct solutions of the wave equations for the fermions in the presence of the particular vortex configurations.

In the same way one may consider the topologically stable $U(1)_B$ vortexes. Then the topological invariant of Eq. (\ref{nuGHall22}) already does not protect the number of the massless fermions localized at these vortexes (its expression is equal to zero). However, one may consider the invariant protected by chiral $Z_2$ symmetry
\begin{eqnarray}
 {\cal N}^\prime_5 =   \frac{2}{5!\,(2\pi)^3}\,\,\int_{} \,{\rm Tr} \Big[ \Gamma^5  \tilde{G}^{(0)} d (\tilde{G}^{(0)})^{-1}\wedge d  \tilde{G}^{(0)} \wedge d(\tilde{G}^{(0)})^{-1} \wedge d  \tilde{G}^{(0)} \wedge d(\tilde{G}^{(0)})^{-1}\Big]  \label{nuGHall228}
\end{eqnarray}
This expression still may be considered as the topological invariant in the approximation when instantons are neglected. It will protect the number of the massless fermions localized on such vortices just like the considered above Eq. (\ref{nuGHall22}).

Thus we come to the conclusion that the approximately stable $U(1)_A$ vortices (and their non - Abelian analogues) carry the exactly massless Majorana fermions, the number of which is given  by the topological invariant of Eq. (\ref{nuGHall22}). At the same time the exactly stable $U(1)_V$ vortices (and their non - Abelian analogues)  carry the approximately massless Majorana fermions, the number of which is given by the topological invariant of Eq. (\ref{nuGHall228}). In the other words, these latter fermions localized on the vortex sheets have masses that appear only thanks to the instantons.  Those masses are supposed to be suppressed as well as the other quantities existing in the CFL phase due to the instantons \cite{Nitta,instantons_CFL}.

\section{The common topological invariant for the gapped phases}

\label{SectGen}

The expression obtained above for the case of the fermions at the background of a vortex Eq. (\ref{nuGHall22}) does not contain the
symmetry matrix specific for the given phase. This prompts that there exists the formulation of the topological invariant in momentum space of the gapped phases, which has the common form for all such phases. We already encountered such a situation in \cite{ZubkovVolovik2012} where the two topological invariants were discussed: ${\cal N}_5$ and ${\cal N}_3$. The ${\cal N}_3$ is the analogue of the symmetry protected topological invariants discussed above while the expression for  ${\cal N}_5$ is similar to that of Eq. (\ref{nuGHall22}). Here we propose the construction similar to that of \cite{ZubkovVolovik2012}, which allows to construct the topological invariants of the same form for all gapped phases of QCD. Namely, we consider the Green function $G$ in the representation of Nambu - Gorkov spinors (where the true chirality matrix is denoted by $\Gamma^5$ rather than $\gamma^5$).  Let us define
\begin{equation}
{\bf G}_{\gamma^5}^{-1} =  e^{i\theta\gamma^5/2} G^{-1} e^{i\theta \gamma^5/2}
\end{equation}
For the zero temperature theory (when the values of $\omega=p^4$ are continuous) we are able to compose out of this modified Green function (depending on extra parameter $\theta$) the following topological invariant:
\begin{eqnarray}
{\cal N}^{\gamma^5}_5 =   \frac{1}{5!\,(2\pi)^3}\,\,\int_{} \,{\rm Tr} \Big[  {\bf G}_{\gamma^5} d ({\bf G}_{\gamma^5})^{-1}\wedge d  {\bf G}_{\gamma^5} \wedge d({\bf G}_{\gamma^5})^{-1} \wedge d  {\bf G}_{\gamma^5} \wedge d({\bf G}_{\gamma^5})^{-1}\Big]  \label{nuGHall229}
\end{eqnarray}
Here the integral is over $S^1\otimes R^4$, where $S^1$ corresponds to $\theta \in [0,2\pi)$.

One can check, that for the non - interacting systems of $n$ Dirac fermions {\it with positive Majorana masses} the expression for this topological invariant is reduced to
\begin{equation}
{\cal N}^{\gamma^5}_5 = n
\end{equation}
The proof of this statement follows the calculation of ${\cal N}_1({\cal C})$ given in Sect. \ref{SectCFLVORT}. It is enough to consider one couple of right - handed and left - handed fermions with equal Majorana masses, Dirac mass $m$, and chemical potential $\mu > m$. (In the absence of $M$ we would have the Fermi surface.) The Green function receives the form
\begin{equation}
\Big[ {\bf G}_{\gamma^5}\Big]^{-1} =  \gamma^\mu p_\mu -  e^{i \gamma^5 \theta } m \gamma^5 \Gamma^0 \Gamma^5 -\mu \gamma^0 \gamma^5 \Gamma^5  - e^{i \gamma^5 \theta } M = e^{i\theta\gamma^5/2}\Big( \gamma^\mu p_\mu - m \gamma^5 \Gamma^0 \Gamma^5 - \mu \gamma^0 \gamma^5\Gamma^5 -  M\Big)e^{i\theta \gamma^5/2}\label{GMm}
\end{equation}
We substitute  this expression to Eq. (\ref{nuGHall229}). Instead of $p_0$ we should substitute into the Green function $i p_4$, where $p_4$ is the imaginary frequency.
The dispersion of quasiparticles is given by
$$
{\cal E} = \pm \sqrt{M^2 + \Big(\mu \pm \sqrt{\vec{p}^2 + m^2}\Big)^2}
$$
Let us suppose first, that the Majorana mass $M$ remains nonzero. Again, the energy gap does not disappear if we smoothly turn off  the chemical potential. It also does not disappear if we turn off Dirac mass $m$. During the decreasing of $\mu$ and $m$ the value of Eq. (\ref{nuGHall229}) is not changed, and we are able to set $\mu = m = 0$ in order to calculate it. The integral over $\theta$ gives the factor $2\pi$ while the integral over the imaginary frequency gives
$$ 2 \int_0^\infty  \frac{d \omega}{(\vec{p}^2+\omega^2+{M}^2)^3}{M}^2 =\frac{3}{8\pi} \frac{{M}^2}{(\vec{p}^2+{M}^2)^{5/2}}$$
We obtain
\begin{eqnarray}
{\cal N}^{\gamma^5}_5 =  - \frac{2\times 5\times 4\times 2\pi  }{5!\,(2\pi)^3}\,\,\int_{} \,{\rm Tr} \Big[ i\gamma^0 \gamma^5 G_0 d G_0^{-1}\wedge d  G_0 \wedge d G_0^{-1}\Big]\times \frac{{M}}{(\vec{p}^2+M^2)^{1/2}} \times \frac{3}{8\pi} \label{nuGHall22992}
\end{eqnarray}
Here the integral is over $3D$ momentum space while
\begin{equation}
{G}_0 =   \gamma^\mu p_\mu  - M
\end{equation}
We have
\begin{eqnarray}
{\cal N}^{\gamma^5}_5 =   \frac{ 1 }{3!\,(2\pi)^2}\,\,\int_{} \,{\rm Tr} \Big[{\bf K}_T \Gamma^5  {G}_0 d ({G}_0)^{-1}\wedge d  {G}_0 \wedge d({G}_0)^{-1}\Big]\times \zeta \label{nuGHall2298}
\end{eqnarray}
where
$$
\zeta = \frac{\int p^2 d p \frac{{M}^2}{(\vec{p}^2+{M}^2)^{5/2}} \times \frac{3}{4\pi}}{\int p^2 dp   \frac{{M}}{(\vec{p}^2+{M}^2)^{2}}} = {\rm sign}\, {M}
$$
This gives
\begin{eqnarray}
{\cal N}_5^{\gamma^5}&=&  1
\end{eqnarray}

By construction Eq. (\ref{nuGHall229}) is well defined in any gapped phase of the theory. Let us calculate it for the one non - interacting fermion with positive Dirac mass. We consider the Green function of the form of Eq. (\ref{GMm}) with $M = \mu = 0$. As a result we come to the following expression for the topological invariant:
\begin{eqnarray}
{\cal N}^{\gamma^5}_5 =   \frac{2\times 5\times 4\times 2\pi  }{5!\,(2\pi)^3}\,\,\int_{} \,{\rm Tr} \Big[ \hat{G}_0 ( -i\gamma^0 \gamma^5 \Gamma^0 \Gamma^5) d \hat{G}_0^{-1}\wedge d  \hat{G}_0 \wedge d \hat{G}_0^{-1}\Big]\times \frac{{m}}{(\vec{p}^2+m^2)^{1/2}} \times \frac{3}{8\pi} \label{nuGHall22993}
\end{eqnarray}
Here the integral is over $3D$ momentum space while
\begin{equation}
\hat{G}_0 =   \gamma^\mu p_\mu  - m \gamma^5 \Gamma^0 \Gamma^5
\end{equation}
Thus we obtain ${\cal N}^{\gamma^5}_5 = 1$ for one massive Dirac fermion. In the case of several massive noninteracting Dirac fermions with positive masses we get
\begin{equation}
{\cal N}^{\gamma^5}_5 = {\cal N}^{CT}_3
\end{equation}

Thus the topological invariant ${\cal N}_5^{\gamma^5}$ defined above has the same value in the CFL phase and in the domains of the HP or QGP, where the theory is connected continuously with the system of non - interacting $u,d,s$ fermions with Dirac masses. This property may be illustrated by the consideration of the interphase between those two phases of quark matter. Let us consider the dispersion of the fermions corresponding to the simplified Green function of the form
\begin{equation}
G^{-1} =  \gamma^\mu p_\mu - m \gamma^5 \Gamma^0 \Gamma^5 + \mu \gamma^0 \gamma^5 \Gamma^5  -  {\bf E} M \label{GMm2}
\end{equation}
It is given by
\begin{equation}
{\cal E} = \pm \sqrt{M^2 + \Big(\mu \pm \sqrt{\vec{p}^2 + m^2}\Big)^2}, \, \pm \sqrt{4 M^2 + \Big(\mu \pm \sqrt{\vec{p}^2 + m^2}\Big)^2}\label{E2}
\end{equation}
This dispersion describes the case when the phase of massive Dirac fermions may coexist with the CFL phase. If $M \ne 0$ we are inside the CFL phase. Let us consider first the case, when $\mu > m$ inside this phase. Then if $M$ is decreased and reaches the value $M=0$, the system drops to the phase with the Fermi surface. The further decreasing of $\mu$ brings us to the gapped phase with the Dirac masses at $\mu < m$. In this pattern we pass through the gapless phase while going between the two different gapped phases. There is another choice of parameters, when from the very beginning in the CFL phase $\mu < m$. Then it does not follow from Eq. (\ref{E2}) that at the interphase between the two gapped phases the massless fermions appear. When $M$ is decreased and reaches the value $M=0$ the CFL phase is transformed smoothly to the gapped phase with Dirac masses of fermions. There will be no such possibility if the values of ${\cal N}_5^{\gamma_5}$ would be different in the mentioned two gapped phases.

\section{Conclusions}

\label{SectConcl}

In this paper we discussed the topological invariants in momentum space in quantum chromodynamics with three flavors of quarks and their relevance for the description of physical phenomena.  Below we summarize our results, enumerate the considered topological invariants and mention their relation to the potentially observed phenomena.

\begin{enumerate}

 \item{} ${\cal N}_3^{Z_2}$ is defined in the quark - gluon plasma phase in the approximation of vanishing current $u$ and $d$ quark masses. It is given by Eq. (\ref{GeneratingFunctionWYC}). For the definition of this invariant we need the continuation of the Green function from the discrete values of imaginary frequencies to the continuous values. This invariant protects the pole (or zero) of this analytically continued fermion propagator corresponding to $u$ and $d$ quarks. The HTL (hard thermal loop) approximation predicts the appearance of the thermal masses of quarks, at least, at asymptotically large temperature. However, instead of the pole the zero of the Green function appears, and we demonstrate, that the value of  ${\cal N}_3^{Z_2} $ is nonzero. When the interactions are turned off, there is the pole of the Green function, but the value of ${\cal N}_3^{Z_2}$ is the same.

\item{} ${\cal N}_5^{\gamma^5}$  is defined in any gapped phase of QCD at vanishing temperature. It is given by Eq. (\ref{nuGHall229}).
${\cal N}^{CT}_3$ is defined in hadronic phase at vanishing chemical potential. It is given by Eq. (\ref{TopInvariantMassive}).  If the system may be reduced by continuous deformation to the noninteracting fermion system with positive masses (which is the most reasonable possibility for the hadronic phase), then ${\cal N}^{CT}_3 = {\cal N}_5^{\gamma^5}$. Thus ${\cal N}_5^{\gamma^5}$ represents the extension of the definition of  ${\cal N}^{CT}_3$ to the general case of the gapped phases of QCD.

    In general case the phase diagram of QCD may contain the axes: temperature, baryonic chemical potential, external magnetic field, external electric potential, angular velocity of rotation, the isotopic and strangeness chemical potentials, etc. As it was mentioned above, we may also add an axis corresponding to the value of $\alpha_s$ at the certain particular scale.
    Thus we describe quark matter that may be hot, dense, may experience the external magnetic field and external electric potential, may be rotated, etc. In spite of the efforts made during the last years in order to investigate the phase structure of QCD \cite{QCDphases,1,2,3,4,5,6,7,8,9,10,CFL0,CFL,m1,m1m8,Nitta,latticeQCD,diquarks,Zhitnitsky}, the behavior of quark matter in many regions of the mentioned above extended phase diagram is not known.

    We point out the hypothetical possibility that somewhere in this phase diagram the topological phase transitions may occur between the sub-phases that differ by the values of the topological invariant ${\cal N}^{CT}_3$. This may occur due to the strong nature of $SU(3)$ interaction between quarks in these phases. The detailed investigation of this possibility is out of the scope of the present paper and it will most likely require the extensive lattice numerical simulations.

\item{} In the hypothetical CFL phase at zero temperature  ${\cal N}_5^{\gamma^5} = N_c N_f = 9$ if the system may be reduced by continuous deformation to the noninteracting fermion system with Majorana masses  (which is almost the only real possibility for the CFL phase).

    According to \cite{Nitta} vortices may appear in rotated quark matter being in the CFL phase. In particular, this may occur in the cores of the dense stars.
    The number of chiral fermions on vortices in the CFL phase is described by the specific topological invariant existing in the mixed $r-p$ space, which is proportional to the product of the vortex topological number $n$ and the value of ${\cal N}^{\gamma^5}_5$.  For the considered non - Abelian vortexes the number of fermion zero modes is $N_c=3$ times smaller than for the Abelian vortexes.

    Massless fermions always invent the important new physics to the description of the real processes. Thus we expect, that the investigation of those objects may become important for the description of  quark matter at extreme conditions. It is worth mentioning, that the topological properties of the other color superconductor phases (like 2SC phase, for example) are in many aspects similar to the considered here CFL phase and may be considered on the same grounds.

\end{enumerate}

To conclude, momentum space topology being applied to QCD allows to investigate the stability of vacuum, it helps in the systematization of the potentially existing phases, and allows to describe successfully the fermion zero modes existing at the topological defects (vortices, etc) and at the surfaces that separate various phases of quark matter.

\section*{Acknowledgements}

The author kindly acknowledges useful discussions with G.E.Volovik, M.N.Chernodub, S.Nikolis, and Z.V.Khaidukov. He is grateful for the kind hospitality at the initial stage of the work to the Laboratory for Mathematics and Theoretical Physics at the University of Tours, France and Le Studium Institute of Advanced Studies.


\begin{thebibliography}{120}

\bibitem{QCDphases}
K.~Fukushima and T.~Hatsuda,
  ``The phase diagram of dense QCD,''
  Rept. Prog. Phys.  {74} (2011) 014001
  doi:10.1088/0034-4885/74/1/014001
  [arXiv:1005.4814 [hep-ph]].

\bibitem{1} A. V. Smilga, “Physics of thermal QCD,” Phys. Rept. 291 (1997) 1–106, arXiv:hep-ph/9612347.

\bibitem{2} K. Rajagopal and F. Wilczek, “The condensed matter physics of QCD,” arXiv:hep-ph/0011333.

\bibitem{3} D. H. Rischke, “The quark-gluon plasma in equilibrium,” Prog. Part. Nucl. Phys. 52 (2004) 197–296, arXiv:nucl-th/0305030

\bibitem{4} M. G. Alford, A. Schmitt, K. Rajagopal, and T. Schafer, “Color superconductivity in dense quark matter,” Rev. Mod. Phys. 80 (2008) 1455–1515, arXiv:0709.4635 [hep-ph]

\bibitem{5} R. S. Hayano and T. Hatsuda, “Hadron properties in the nuclear medium,” arXiv:0812.1702 [nucl-ex]. [17] H. Satz, “The states of matter in QCD,” arXiv:0903.2778 [hep-ph].

\bibitem{6}  M. Huang, “QCD phase diagram at high temperature and density,” arXiv:1001.3216 [hep-ph]. [19] A. Schmitt, “Dense matter in compact stars - A pedagogical introduction,” arXiv:1001.3294 [astro-ph.SR].

\bibitem{7} J.~O.~Andersen, W.~R.~Naylor and A.~Tranberg,
  ``Phase diagram of QCD in a magnetic field: A review,''
  Rev.\ Mod.\ Phys.\  {\bf 88} (2016) 025001
  doi:10.1103/RevModPhys.88.025001
  [arXiv:1411.7176 [hep-ph]].

\bibitem{8} K.~Fukushima and C.~Sasaki,
  ``The phase diagram of nuclear and quark matter at high baryon density,''
  Prog.\ Part.\ Nucl.\ Phys.\  {\bf 72} (2013) 99
  doi:10.1016/j.ppnp.2013.05.003
  [arXiv:1301.6377 [hep-ph]].

\bibitem{9} V.~A.~Miransky and I.~A.~Shovkovy,
  ``Quantum field theory in a magnetic field: From quantum chromodynamics to graphene and Dirac semimetals,''
  Phys.\ Rept.\  {\bf 576} (2015) 1
  doi:10.1016/j.physrep.2015.02.003
  [arXiv:1503.00732 [hep-ph]].

\bibitem{10} A.~R.~Zhitnitsky,
  ``QCD as a topologically ordered system,''
  Annals Phys.\  {\bf 336} (2013) 462
  doi:10.1016/j.aop.2013.05.020
  [arXiv:1301.7072 [hep-ph]].

\bibitem{CFL0}
Mark Alford, Krishna Rajagopal, Frank Wilczek,
Nuclear Physics B {\bf 537}, 443--458 (1999).

\bibitem{CFL}
Michael Buballa, Phys.Rept. {\bf 407},  205--376 (2005).

\bibitem{m1}
Roberto Anglani, Massimo Mannarelli, Marco Ruggieri,
"Collective modes in the color flavor-locked phase",
New J. Phys.{\bf 13}, 055002 (2011).

\bibitem{m1m8}
Minoru Eto, Muneto Nitta, and Naoki Yamamoto, PRL {\bf 104}, 161601 (2010)

Minoru Eto and Muneto Nitta, Phys. Rev. D {\bf 80}, 125007 (2009)

\bibitem{Nitta}
  M.~Eto, Y.~Hirono, M.~Nitta and S.~Yasui,
 ``Vortices and Other Topological Solitons in Dense Quark Matter,''
  PTEP {\bf 2014} (2014) no.1,  012D01
  doi:10.1093/ptep/ptt095
  [arXiv:1308.1535 [hep-ph]].


\bibitem{latticeQCD}
 C. DeTar and U. M. Heller, “QCD thermodynamics from the lattice,” Eur. Phys. J. A41 (2009) 405–437, arXiv:0905.2949 [hep-lat].

\bibitem{diquarks}
D. Ebert, K.G. Klimenko, V.L. Yudichev,
Eur. Phys. J. C {\bf 53}, 65--76 (2008).

\bibitem{Zhitnitsky}
  M.~M.~Forbes and A.~R.~Zhitnitsky,
  ``Domain walls in QCD,''
  JHEP {\bf 0110} (2001) 013
  doi:10.1088/1126-6708/2001/10/013
  [hep-ph/0008315].


\bibitem{Volovik2003}
G.E. Volovik,
{\it The Universe in a Helium Droplet},
Clarendon Press,  Oxford (2003).


\bibitem{So1985}
H. So, Induced topological invariants by lattice fermions in odd dimensions,
Prog. Theor. Phys. {\bf 74}, 585--593 (1985).

\bibitem{IshikawaMatsuyama1986}
K. Ishikawa  and T. Matsuyama, Magnetic field induced multi component QED in
three-dimensions and quantum Hall effect, Z. Phys. C {\bf 33}, 41--45 (1986).

\bibitem{Kaplan1992}
 D.B. Kaplan,
Method for simulating chiral fermions on the lattice, Phys. Lett.  B {\bf 288},
342--347 (1992); arXiv:hep-lat/9206013.

\bibitem{Golterman1993}
 M.F.L. Golterman, K.  Jansen and D.B. Kaplan,
Chern-Simons  currents and chiral  fermions on the lattice,
 Phys.Lett. B {\bf 301}, 219--223 (1993):
arXiv: hep-lat/9209003.

\bibitem{Horava2005}
P. Ho\v{r}ava, Stability of Fermi surfaces and $K$-theory, Phys. Rev. Lett.
\textbf{95}, 016405 (2005).

\bibitem{Creutz2008}
M. Creutz, Four-dimensional graphene and chiral fermions, JHEP 04 (2008) 017;
arXiv:0712.1201.

\bibitem{Kaplan2011}
D.B. Kaplan and Sichun Sun, Spacetime as a topological insulator,
arXiv:1112.0302.

\bibitem{Creutz2011}
M. Creutz,
Confinement, chiral symmetry, and the lattice,
arXiv:1103.3304.


  \bibitem{Volovik2000}
  G.E. Volovik,
  Momentum-space topology of Standard Model,
  J. Low Temp. Phys.,  {\bf 119}, 241 -- 247 (2000); hep-ph/9907456.

  \bibitem{Volovik2010}
G.E. Volovik,
 Topological invariants  for Standard Model: from semi-metal to topological insulator,
 Pis'ma ZhETF {\bf 91}, 61--67 (2010);   JETP Lett. {\bf 91}, 55--61 (2010);
arXiv:0912.0502.

\bibitem{ZubkovVolovik2012}
M.A. Zubkov, G.E. Volovik
Momentum space topological invariants for the 4D relativistic vacua with mass gap,
 Nucl. Phys. B {\bf 860}, 295--309  (2012) .

\bibitem{Zubkov2012}
M.A.Zubkov,
Momentum space topology in the lattice gauge theory,
 arXiv:1210.1989.

 \bibitem{VZ2016}
   G.~E.~Volovik and M.~A.~Zubkov,
  ``Standard Model as the topological material,''
  New J.\ Phys.\  {\bf 19} (2017) no.1,  015009
  doi:10.1088/1367-2630/aa573d
  [arXiv:1608.07777 [hep-ph]].

\bibitem{Z2016}
  M.~A.~Zubkov,
  ``Absence of equilibrium chiral magnetic effect,''
  Phys.\ Rev.\ D {\bf 93} (2016) no.10,  105036
  doi:10.1103/PhysRevD.93.105036
  [arXiv:1605.08724 [hep-ph]].

\bibitem{Z2016AOP}
  M.~A.~Zubkov,
  ``Wigner transformation, momentum space topology, and anomalous transport,''
  Annals Phys.\  {\bf 373} (2016) 298
  doi:10.1016/j.aop.2016.07.011
  [arXiv:1603.03665 [cond-mat.mes-hall]].

\bibitem{HasanKane2010}
M.Z. Hasan and C.L. Kane,
Topological Insulators,
Rev. Mod. Phys. {\bf 82}, 3045--3067 (2010).

\bibitem{Xiao-LiangQi2011}
Xiao-Liang Qi and Shou-Cheng Zhang,
Topological insulators and superconductors,
Rev. Mod. Phys. {\bf 83}, 1057--1110 (2011).

\bibitem{Schnyder2008}
A.P. Schnyder, S. Ryu, A. Furusaki and A.W.W. Ludwig,
Classification of topological insulators and superconductors in three spatial dimensions,
Phys. Rev. {\bf B~ 78}, 195125 (2008); A.P. Schnyder, S. Ryu, A. Furusaki and A.W.W. Ludwig,
Classification of topological insulators and superconductors,
 AIP Conf. Proc. {\bf 1134}, 10 (2009);
 arXiv:0905.2029.

\bibitem{Kitaev2009}
A. Kitaev,
Periodic table for topological insulators and superconductors,
AIP Conference Proceedings, Volume {\bf 1134}, pp. 22--30 (2009);
  arXiv:0901.2686.

\bibitem{Grinevich1988}
 P.G. Grinevich, G.E. Volovik,
 Topology of gap nodes in superfluid $^3$He: $\pi_4$ homotopy group for $^3He-B$
  disclination, J. Low Temp. Phys. {\bf 72}, 371--380  (1988).


\bibitem{BevanNature1997}
 T.D.C. Bevan, A.J. Manninen, J.B. Cook, J.R. Hook et al.,
Momentogenesis by $^3$He vortices: an experimental  analog of primordial baryogenesis,
Nature,  {\bf 386},  689--692 (1997).



\bibitem{Soluyanov2015}
A.A. Soluyanov, D. Gresch, Zhijun Wang, QuanSheng Wu, M. Troyer, Xi Dai, B.A. Bernevig,
Type-II Weyl Semimetals,
 Nature {\bf 527}, 495--498 (2015).

\bibitem{SchnyderBrydon2015}
A.P. Schnyder, P.M.R. Brydon,
Topological surface states in nodal superconductors,
J. Phys.: Condens. Matter {\bf 27}, 243201 (2015).

\bibitem{Mizushima2016}
T. Mizushima, Ya. Tsutsumi, T. Kawakami, M. Sato, M. Ichioka, K. Machida,
Symmetry protected topological superfluids and superconductors - From the basics to $^3$He,
J. Phys. Soc. Jpn. {\bf 85}, 022001 (2016).

\bibitem{Bansil2016}
A. Bansil, Hsin Lin, Tanmoy Das,
Colloquium:  Topological band theory,
Rev. Mod. Phys.{\bf 88}, 021004 (2016).

\bibitem{Yonezawa2016}
Shingo Yonezawa,
Bulk topological superconductors,
 AAPPS Bulletin, {\bf 26},  3 (2016).

\bibitem{Witten2015}
Edward Witten,
Fermion Path Integrals And Topological Phases,
arXiv:1508.04715.




\bibitem{Wang2014}
Chong Wang, T. Senthil,
Interacting fermionic topological insulators/superconductors in three dimensions,
Phys. Rev. B {\bf 89}, 195124 (2014).


\bibitem{Kapustin2015}
Anton Kapustin, Ryan Thorngren, Alex Turzillo, Zitao Wang,
Fermionic Symmetry Protected Topological Phases and Cobordisms,
JHEP 1512:052,2015, arXiv:1406.7329



\bibitem{Kitaev}
Alexei Kitaev,
Homotopy-theoretic approach to SPT phases in action: $Z_{16}$ classification of three-dimensional superconductors,
http://www.ipam.ucla.edu/abstract/?tid=12389\&pcode=STQ2015


 \bibitem{Gurarie2011}
 V. Gurarie,
 Single-particle Green-s functions and interacting topological insulators,
 Phys. Rev. B {\bf 83}, 085426 (2011).

 \bibitem{EssinGurarie2011}
A.M. Essin and V. Gurarie,
Bulk-boundary correspondence of topological insulators from their Green's functions,
Phys. Rev. B {\bf 84}, 125132 (2011).




\bibitem{Volovik2009}
 G.E. Volovik,
"Fermion zero modes at the boundary of superfluid $^3$He-B,"
 Pis'ma ZhETF {\bf 90}, 440--442 (2009); JETP Lett. {\bf 90}, 398--401 (2009);
arXiv:0907.5389;
"Topological invariant  for superfluid  $^3$He-B and quantum phase transitions",
Pis'ma ZhETF {\bf 90}, 639--643 (2009);  JETP Lett. {\bf 90}, 587--591 (2009);
arXiv:0909.3084;
"Topological superfluid $^3$He-B in magnetic field and Ising variable",
 Pis'ma ZhETF {\bf 91},  215--219 (2010);  JETP Lett. {\bf 91}, 201--205 (2010);
arXiv:1001.1514.

\bibitem{Silaev}
  M.~A.~Silaev and G.~E.~Volovik,
  ``Topological superfluid $^3$He-B: Fermion zero modes on interfaces and in the vortex core,''
  J.\ Low.\ Temp.\ Phys.\  {\bf 161} (2010) 460
  doi:10.1007/s10909-010-0226-z
  [arXiv:1005.4672 [cond-mat.other]].


\bibitem{heavyions}
 N. Armesto, (ed. ) et al., “Heavy Ion Collisions at the LHC - Last Call for Predictions,” J. Phys. G35 (2008) 054001, arXiv:0711.0974 [hep-ph].

 \bibitem{stars}  A. Schmitt, “Dense matter in compact stars - A pedagogical introduction,” arXiv:1001.3294 [astro-ph.SR].

\bibitem{Nishida:2010wr}
  Y.~Nishida,
  ``Is a color superconductor topological?,''
  Phys.\ Rev.\ D {\bf 81} (2010) 074004
  doi:10.1103/PhysRevD.81.074004
  [arXiv:1001.2555 [hep-ph]].

\bibitem{nitta_1}
  S.~Yasui, K.~Itakura and M.~Nitta,
  ``Fermion structure of non-Abelian vortices in high density QCD,''
  Phys.\ Rev.\ D {\bf 81} (2010) 105003
  doi:10.1103/PhysRevD.81.105003
  [arXiv:1001.3730 [math-ph]].

\bibitem{nitta_2}
  T.~Fujiwara, T.~Fukui, M.~Nitta and S.~Yasui,
  ``Index theorem and Majorana zero modes along a non-Abelian vortex in a color superconductor,''
  Phys.\ Rev.\ D {\bf 84} (2011) 076002
  doi:10.1103/PhysRevD.84.076002
  [arXiv:1105.2115 [hep-ph]].
  






















\bibitem{Wigner}  E.P. Wigner, "On the quantum correction for thermodynamic equilibrium", Phys. Rev. 40 (June 1932) 749–759. doi:10.1103/PhysRev.40.749


\bibitem{star2} 	
F. Bayen, M. Flato, C. Fronsdal, A. Lichnerowicz, D. Sternheimer, "Deformation Theory and Quantization. 2. Physical Applications",
 Annals Phys. 111 (1978) 111


\bibitem{Weyl}
Robert G Littlejohn, "The semiclassical evolution of wave packets",
Physics Reports Volume 138, Issues 4–5, May 1986, Pages 193-291

\bibitem{berezin} Berezin, F.A. and M.A. Shubin, 1972, in: Colloquia Mathematica Societatis Janos Bolyai (North-Holland, Amsterdam) p. 21.

\bibitem{vortex}
 Kei Iida and Gordon Baym, Phys.Rev., D66, 014015 (2002)

 Michael McNeil Forbes and Ariel R. Zhitnitsky, Phys.Rev., D65, 085009 (2002).

\bibitem{sigma_space}
  E.~Laermann {\it et al.} [RBC-Bielefeld Collaboration],
  Nucl.\ Phys.\ A {\bf 820} (2009) 223C.
  doi:10.1016/j.nuclphysa.2009.01.055

\bibitem{instantons_CFL}
  T.~Schafer,
  Phys.\ Rev.\ D {\bf 65} (2002) 094033
  doi:10.1103/PhysRevD.65.094033
  [hep-ph/0201189].

\bibitem{LeBellac}
M. Le Bellac "Thermal Field Theory", (2000) Cambridge University Press, Cambridge.

\bibitem{deForcrand:2017cgb}
  P.~de Forcrand and M.~D'Elia,
  ``Continuum limit and universality of the Columbia plot,''
  PoS LATTICE {\bf 2016} (2017) 081
  [arXiv:1702.00330 [hep-lat]].

\bibitem{Schafer}
  T.~Schäfer,
  ``Patterns of symmetry breaking in QCD at high baryon density,''
  Nucl.\ Phys.\ B {\bf 575} (2000) 269
  doi:10.1016/S0550-3213(00)00063-8
  [hep-ph/9909574].


\bibitem{spinors}
 D. Bailin and A. Love, Supersymmetric Gauge Field Theory and String Theory, (IOP, London, 1994).

\bibitem{Zubkov:2012nm}
  M.~A.~Zubkov,
  ``Generalized unparticles, zeros of the Green function, and momentum space topology of the lattice model with overlap fermions,''
  Phys.\ Rev.\ D {\bf 86} (2012) 034505
  doi:10.1103/PhysRevD.86.034505
  [arXiv:1202.2524 [hep-lat]].

\bibitem{Polykarp}
  M.~I.~Polikarpov,
``Recent results on the Abelian projection of lattice gluodynamics,''
  Nucl.\ Phys.\ Proc.\ Suppl.\  {\bf 53} (1997) 134
  doi:10.1016/S0920-5632(96)00607-X
  [hep-lat/9609020].

\bibitem{NJL}
M.K. Volkov and A. Radzhabov, Phys. Usp. 49, 551 (2006).

\bibitem{liquid}
Schafer, T., Shuryak, E. V., "Instantons in QCD", Reviews of Modern Physics, 70 (1998), 323–426, arXiv:hep-ph/9610451

\bibitem{Born}
  V.~G.~Bornyakov, D.~L.~Boyda, V.~A.~Goy, A.~V.~Molochkov, A.~Nakamura, A.~A.~Nikolaev and V.~I.~Zakharov,
  ``New approach to canonical partition functions computation in $N_f=2$ lattice QCD at finite baryon density,''
  Phys.\ Rev.\ D {\bf 95} (2017) no.9,  094506
  doi:10.1103/PhysRevD.95.094506
  [arXiv:1611.04229 [hep-lat]].

\end{thebibliography}
\end{document}